\documentclass[times,twocolumn]{aastex62}

\usepackage{CJK}
\usepackage{amsmath}
\usepackage{multirow}
\usepackage{cases}
\usepackage{graphicx}
\usepackage{subfigure}
\usepackage{natbib}
\usepackage{color}
\usepackage{array}
\usepackage{tabularx}
\usepackage{bm}
\usepackage{threeparttable}
\usepackage{gensymb}

\defcitealias{CCM89}{CCM89}

\DeclareGraphicsExtensions{.pdf,.png,.jpg}

\shorttitle{}
\shortauthors{Zhang et al.}

\begin{document}
\begin{CJK*}{UTF8}{gbsn}


\title{{\large Enabling Metallicity Measurements of M-dwarf Microlensing Lenses out to the Galactic Bulge}}






\correspondingauthor{Jiyuan Zhang}
\email{zhangjiyuan2022@gmail.com}

\author[0000-0002-1279-0666]{Jiyuan Zhang$^*$}
\affiliation{Department of Astronomy, Tsinghua University, Beijing 100084, China}
\affiliation{Department of Astronomy, Westlake University, Hangzhou 310030, Zhejiang Province, China}

\author[0000-0001-5417-2260]{Gregory Green}
\affiliation{Department of Astronomy, Westlake University, Hangzhou 310030, Zhejiang Province, China}

\author[0000-0001-6000-3463]{Weicheng Zang}
\affiliation{Department of Astronomy, Westlake University, Hangzhou 310030, Zhejiang Province, China}

\author[0000-0003-0626-8465]{Hongjing Yang}
\affiliation{Department of Astronomy, Westlake University, Hangzhou 310030, Zhejiang Province, China}

\author[0000-0001-8317-2788]{Shude Mao}
\affiliation{Department of Astronomy, Westlake University, Hangzhou 310030, Zhejiang Province, China}

\begin{abstract}
We demonstrate that metallicity becomes identifiable for M-dwarf microlensing lenses out to bulge distances when multi-band lens photometry is combined with the angular Einstein radius, thereby defining the photometric--microlensing metallicity method.
The cool atmospheres of M dwarfs contain abundant molecules, whose broad absorption bands make their positions in an optical--NIR color--absolute magnitude diagram sensitive to metallicity. 
The angular Einstein radius provides the mass--distance constraint needed to infer the lens absolute magnitude, while the intrinsic M-dwarf locus and reddening vector are non-parallel in color--color space, allowing the lens extinction and intrinsic color to be inferred simultaneously from its multi-band photometry, thereby recovering the lens intrinsic position in the metallicity-sensitive color--absolute magnitude diagram. 
The method requires three-band lens photometry spanning roughly the \(R\), \(Z\), and \(K\) bands, making it particularly well suited to Roman through its high-resolution $F062$, $F087$, and $F213$ imaging. 
With the planned Roman Galactic Bulge Time-Domain Survey (GBTDS) observations plus an additional \(\sim8~\mathrm{hr}\) of \(F062\) imaging per field (\(\sim40~\mathrm{hr}\) in total) roughly a decade after Roman launch, host metallicities could be measured for \(\sim150\) planetary systems under the GBTDS yield forecast, with a typical \(1\sigma\) precision of \(\sim0.25\) dex from our mock-recovery analysis. 
Applied homogeneously to GBTDS microlensing events with and without detected planets, the method would enable the first measurement of the occurrence--metallicity relation for cold low-mass planets beyond the snow line, constrain the low-metallicity cutoff for their formation, and extend occurrence--metallicity studies into the inner Galaxy. 
\end{abstract}

\section{Introduction}\label{sec:1_intro}

Host-star metallicity traces the metal content of the protoplanetary disk and hence the amount of solid material available for planet formation. 
Planet occurrence as a function of host metallicity therefore directly probes how the available solid material affects planet formation. 
The planet occurrence--host metallicity relation shows a clear dependence on planet mass. 
Giant-planet occurrence rises steeply with host metallicity \citep{Santos2004,FischerValenti2005,Neves2013}, while lower-mass planets show a weaker relation, with sub-Neptunes still showing a positive correlation and super-Earths showing a nearly flat relation \citep{Sousa2011,Petigura2018,Lu2020}, a trend seen around both FGK and M dwarfs. 
This mass dependence arises naturally in core-accretion models, in which higher solid content allows cores to grow rapidly enough to reach the critical mass for runaway gas accretion before disk dispersal, whereas lower-mass planets can form without reaching this threshold, making their formation less sensitive to metallicity \citep{Pollack1996,IdaLin2004b,Mordasini2012}. 

As planet samples have grown, an additional dependence on orbital period has emerged. 
For giant planets, hot-Jupiter hosts tend to be more metal-rich than cold-Jupiter hosts \citep{Adibekyan2013,Maldonado2018,Buchhave2018,Gan2025}. 
In the standard core-accretion scenario, giant-planet formation is favored beyond the water snow line, where water vapor condenses into ice and increases the surface density of solids, while hot Jupiters subsequently migrate inward to short-period orbits \citep{Pollack1996,Lin1996}. 
The metallicity difference between hot- and cold-Jupiter hosts may therefore point to metallicity-dependent migration histories, either through planet--disk interactions, as faster giant-planet growth in metal-rich disks leaves more time for inward migration before disk dispersal, or through planet--planet scattering, which is expected to occur more frequently in metal-rich systems owing to their greater tendency to form multiple giant planets \citep{ColemanNelson2016,DawsonMurrayClay2013}. 
A similar period dependence is seen for smaller planets, with a positive occurrence--metallicity relation for hot super-Earths with periods of 1--10 days but a nearly flat relation for warm super-Earths with periods of 10--100 days, and with a steeper relation for hot than warm sub-Neptunes \citep{Petigura2018}. 
Within the parameter space currently probed, the occurrence--metallicity relation therefore tends to steepen toward larger planets and shorter orbital periods. 
However, how this trend extends to cold low-mass planets beyond the snow line, which constitute the currently missing quadrant of occurrence--metallicity measurements, remains unclear, as such planets are largely inaccessible to transit, radial-velocity, and astrometric surveys. 

Despite the weak dependence of low-mass planet occurrence on host metallicity, low-mass planet formation is expected to eventually be suppressed at sufficiently low metallicity. 
Such a low-metallicity cutoff can arise naturally because lower metallicity reduces the solid reservoir in protoplanetary disks, eventually leaving insufficient solid material to form even low-mass planets. 
At lower stellar masses, the smaller disk mass is expected to shift the cutoff to higher metallicity \citep{Lu2020}. 
More broadly, measuring the metallicity at which this cutoff occurs would constrain the minimum solid reservoir required to form low-mass planets and the efficiency with which disk solids are assembled into planets, making the cutoff a sensitive probe of planet-formation models. 
An indication of such a cutoff has recently been reported for hot super-Earths with periods of 1--10 days around FGK stars, with a sharp drop in occurrence in the \(-0.75<[\mathrm{Fe/H}]\leq-0.5\) bin \citep{Boley2024}. 
However, the metallicity at which this cutoff occurs for cold low-mass planets remains entirely unconstrained. 

More generally, existing occurrence--metallicity measurements are dominated by samples from the solar neighborhood and the Kepler field, while the relation in the inner Galaxy remains largely unprobed. 
Measuring host metallicities in the inner Galaxy would therefore test whether the occurrence--metallicity relation established locally also applies in a different Galactic environment. 
Moreover, host metallicities are required for a clean comparison of planet occurrence across Galactic environments, because without this information an apparent environmental dependence cannot be disentangled from a metallicity dependence associated with differences in the underlying metallicity distributions \citep{Penny2016,Zink2023}. 
The Galactic bulge offers an additional advantage through its broad metallicity range. 
Compared with the relatively narrow solar-neighborhood metallicity distribution with a mean near solar metallicity and a standard deviation of $\sim0.2$ dex \citep{Casagrande2011}, the bulge has a much broader, multi-component metallicity distribution. 
Even a bulge field near the planned Roman microlensing footprint but covering a much smaller area already shows substantial probability density over $-0.6\lesssim[\mathrm{Fe/H}]\lesssim+0.4$, with a metal-poor tail extending to $[\mathrm{Fe/H}]\sim-1$ \citep{Zoccali2017_bulge_metallicity}.
This provides a much broader metallicity baseline for measuring the occurrence--metallicity relation, while the metal-poor tail provides particular leverage on constraining the low-metallicity cutoff for cold low-mass planet formation. 

Gravitational microlensing is uniquely sensitive to planets over a wide range of masses near and beyond the snow line, around host stars spanning the Galactic disk and bulge \citep{MaoPaczynski1991,GouldLoeb1992,BennettRhie1996,Gaudi2012}. 
For cold low-mass planets in particular, microlensing is expected to remain the only probe in the foreseeable future. 
Microlensing observations indicate that super-Earths on Jupiter-like orbits are common \citep{Zang2025}. 
The Galactic Bulge Time-Domain Survey (GBTDS) of the Nancy Grace Roman Space Telescope is expected to detect approximately \(1400\) bound planets through microlensing, including roughly \(200\) with masses below \(3\,M_\oplus\) \citep{Penny2019}. 
With host metallicity measurements, microlensing planets would enable the first measurement of the occurrence--metallicity relation for cold low-mass planets and extend occurrence--metallicity studies from the solar neighborhood to the inner Galaxy. 
However, metallicity measurements are generally unavailable for microlensing lenses because they are typically faint, distant, and blended with the source star. 
Simulations predict that most Roman lenses will lie in the Galactic bulge, with the lens-distance distribution peaking near \(7.5\,\mathrm{kpc}\), and that the lens population will be dominated by M dwarfs \citep{Penny2019,paper4_Terry2026}. 
To obtain a sufficiently large sample for occurrence--metallicity studies, a lens-metallicity method must therefore be applicable to M-dwarf lenses out to bulge distances. 

In this work, we develop the photometric--microlensing metallicity method, which combines multi-band lens photometry with the angular Einstein radius  to measure the metallicities of M-dwarf lenses out to bulge distances. 
The method builds on the established M-dwarf photometric metallicity method: the cool atmospheres of M dwarfs contain abundant molecules whose broad absorption bands make optical--near infrared (NIR) colors strongly metallicity dependent at fixed near-infrared absolute magnitude \citep[e.g.,][]{Bonfils2005,Neves2012}. 
To extend this approach to microlensing lenses, which generally lack astrometric parallaxes and suffer strong reddening, the angular Einstein radius provides the mass--distance constraint needed to infer the lens absolute magnitude, while the non-parallel orientations of the M-dwarf locus and reddening vector in color--color space enable the lens to be dereddened from its multi-band photometry. 
The method thereby recovers the intrinsic position of the lens on the metallicity-sensitive color--absolute magnitude diagram and jointly constrains its mass, distance, extinction, and metallicity. 

The method is particularly well suited to Roman, which will concentrate a large microlensing sample within a compact survey footprint through its GBTDS, and can provide the high-resolution imaging required for multi-band lens photometry. 
The method requires three-band photometry spanning an optical--NIR wavelength range roughly corresponding to the \(R\), \(Z\), and \(K\) bands, realized by the $F062$, $F087$, and $F213$ filters of the Wide Field Instrument (WFI) in the Roman implementation. 
With the currently planned GBTDS observations plus an additional \(\sim8~\mathrm{hr}\) of \(F062\) imaging per field roughly a decade after Roman launch, corresponding to a total of \(\sim40~\mathrm{hr}\) to cover all five contiguous GBTDS fields, we estimate that host metallicities could be measured for \(\sim150\) planetary systems under the \citet{Penny2019} yield forecast, with a typical \(1\sigma\) precision of \(\sim0.25\) dex from our mock-recovery analysis. 
Because the method relies on wide-field photometry rather than target-by-target spectroscopy, it can also be applied to eligible events without detected planets at no additional observing cost, enabling homogeneous metallicity inference for both these lenses and the planet hosts and thereby providing the parent sample required for the occurrence--metallicity analysis.

The paper is organized as follows.
Section~\ref{sec:2_principle} presents the principle of the photometric--microlensing metallicity method. 
Section~\ref{sec:3_mock_recovery_experiments} assesses its expected performance through mock-recovery experiments.
Section~\ref{sec:4_discussion} discusses empirical calibration, applicability, observational requirements, parent-sample construction, unresolved companions, and spectroscopic validation. 
All magnitudes in this paper are given in the Vega system. 







\section{Principle of the Method}\label{sec:2_principle}

In this section, we present the principle of the photometric--microlensing metallicity method. 
We first show that optical--NIR color--absolute magnitude diagram (CMD) position can serve as a metallicity probe for M dwarfs. 
We then show how microlensing observables constrain the lens absolute magnitude and how multi-band lens photometry enables color--color dereddening to recover the lens dereddened optical--NIR color, together placing M-dwarf lenses on this CMD for the metallicity inference. 
Finally, we describe the calculation and empirical calibration of the reddening-vector direction required for the dereddening. 

\subsection{Optical--NIR CMD position as an M-dwarf metallicity probe}
\label{subsec:2_1}

Measuring M-dwarf metallicities spectroscopically is difficult. 
For FGK stars, metallicities and detailed elemental abundances can often be measured spectroscopically from the equivalent widths of relatively isolated atomic lines \citep{Reddy2003, Sousa2008, Adibekyan2012, Bensby2014_Disk}. 
This approach is much harder for M dwarfs. 
Their cool atmospheres allow abundant diatomic and triatomic molecules to form (e.g., TiO, VO, CaH, MgH, FeH, H$_2$O, and CO), and dense forests of molecular lines blend into broad absorption bands. 
These bands suppress flux over wide wavelength ranges, hiding the true continuum and leaving a depressed pseudo-continuum. 
See \href{https://www.astroexplorer.org/details/apjsaa656df9}{Figure~9} of \citet{Kesseli2017} for a comparison of M-dwarf and hotter-star spectra. 
Because equivalent-width measurements require a well-defined continuum reference, this pseudo-continuum prevents the classical line-based analysis used for FGK stars from being directly extended to M dwarfs. 
Full spectral synthesis provides an alternative, but purely model-based abundance inference remains challenging because synthetic spectra still show significant mismatches to observed M-dwarf spectra, largely due to incomplete or imperfect molecular opacity data for dominant absorbers, including TiO in optical \citep{Hoeijmakers2015_TiO,McKemmish2019_TiO,Rains2021,Rains2024}.

The same molecular absorption bands that complicate M-dwarf spectroscopy also imprint a strong metallicity dependence on broad-band fluxes. 
At fixed mass, higher metallicity increases the atmospheric opacity mainly at optical wavelengths, where TiO and VO bands dominate, suppressing the optical flux and shifting the flux distribution toward the near-infrared. 
At the same time, higher metallicity lowers the bolometric luminosity at fixed mass, because the increased opacity leads to a cooler, less luminous stellar structure. 
These two effects act in the same direction in optical bands, making metal-rich M dwarfs optically fainter at fixed mass. 
In the near-infrared, however, the redward shift of the flux distribution counteracts the lower bolometric luminosity, making near-infrared absolute magnitudes largely insensitive to metallicity \citep{Chabrier_and_Baraffe2000,Delfosse2000,Bonfils2005,Neves2012,Mann2019}. 
Together with the fact that main-sequence M dwarfs evolve negligibly over the age of the universe, this makes a near-infrared absolute magnitude primarily trace stellar mass, while the optical--NIR color at fixed near-infrared absolute magnitude mainly reflects metallicity. 
The position of an M dwarf on an optical--NIR CMD is therefore a sensitive metallicity probe. 

This color--magnitude metallicity dependence can be empirically calibrated using FGK+M binaries. 
Because binary stars form from the same molecular cloud, they are expected to share the same birth composition, an assumption supported by the well-established chemical homogeneity of FGK+FGK binaries \citep{Desidera2004_FGK_FGK_binary,Hawkins2020_FGK_FGK_binary}. 
The metallicity of the FGK primary can be measured spectroscopically and assigned to the M-dwarf companion, yielding an empirical calibration between M-dwarf optical--NIR CMD position and metallicity \citep{Bonfils2005,Johnson2009,Schlaufman2010,Neves2012,Rains2021,Duque-Arribas2023}. 
These empirical calibrations established what are commonly referred to as photometric metallicity methods for M dwarfs. 
Recent calibrations can reach a \(\sim 0.1\) dex scatter \citep{Duque-Arribas2023}, with the residuals likely reflecting intrinsic astrophysical dispersion, such as rotation and magnetic activity \citep{Neves2012,Duque-Arribas2023}. 

\begin{figure}
    \centering
    \includegraphics[width=\columnwidth]{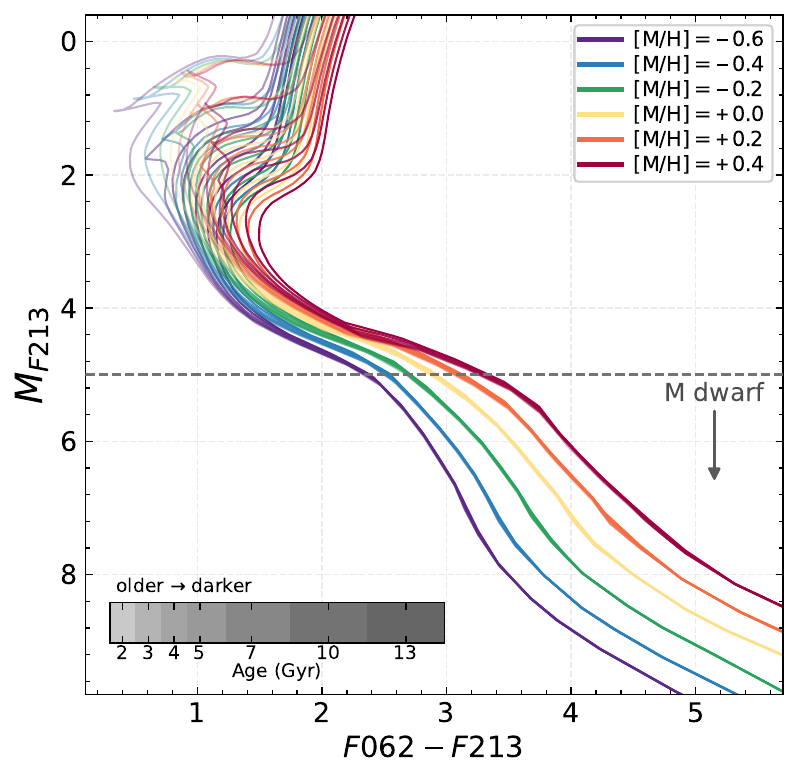}
    \caption{
    Optical--NIR color--absolute magnitude diagram in Roman \(F062\) and \(F213\). 
    The curves show PARSEC v1.2S isochrones, with color indicating metallicity and transparency indicating age. 
    In the M-dwarf regime, where \mbox{\(M_{F213}\gtrsim5\)}~mag, isochrones of different metallicity separate strongly in \(F062-F213\), whereas isochrones of different age largely overlap. 
    At brighter magnitudes, isochrones of different metallicity are less separated along the FGK-dwarf sequence, while both age and metallicity  substantially shift the turnoff and subgiant tracks. 
    Figures~\ref{fig:mass_luminosity_relation} and~\ref{fig:color_color_plot} show how microlensing constraints first recover the lens absolute magnitude and how multi-band lens photometry then recovers the dereddened optical--NIR color, together placing M-dwarf lenses on this CMD for the metallicity inference.
}
    \label{fig:cmd}
\end{figure}

\begin{figure}
    \centering
    \includegraphics[width=\columnwidth]{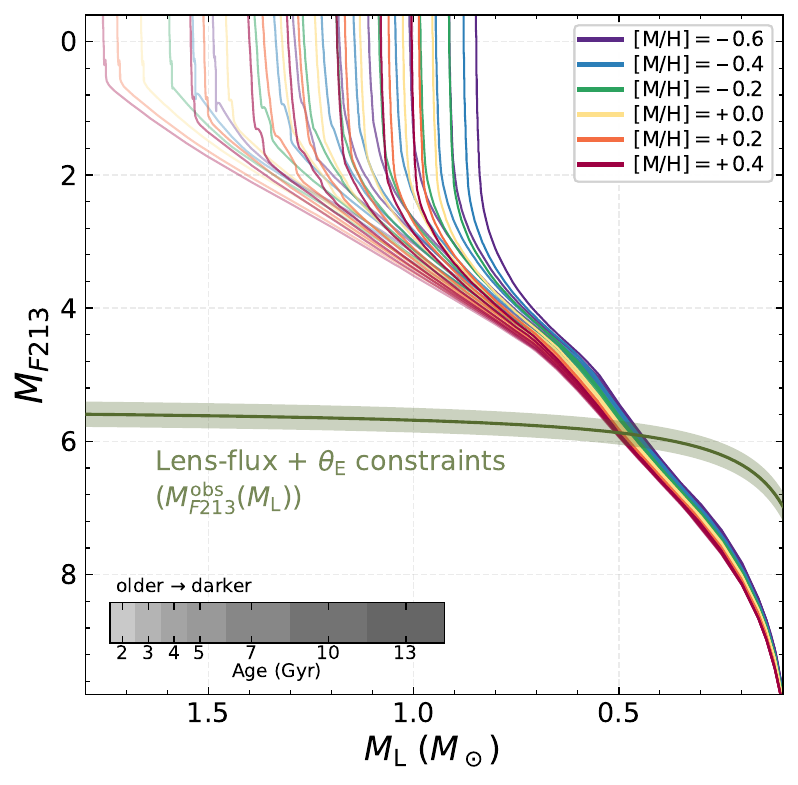}
    \caption{
    Lens-flux plus \(\theta_{\rm E}\) constraints in the \(M_{F213}\)--\(M_{\rm L}\) plane for the fiducial bulge M-dwarf lens event used in the mock-recovery experiment of Section~\ref{sec:3_mock_recovery_experiments}. 
    Among the mass--distance constraints discussed in Section~\ref{subsec:2_2}, this pair is expected to be the most frequently available for Roman microlensing events. 
    The PARSEC v1.2S isochrones show the intrinsic \(M_{F213}\)--\(M_{\rm L}\) relation set by stellar physics, with color indicating metallicity and transparency indicating age. 
    The green band shows \(M_{F213}^{\rm obs}(M_{\rm L})\), which is the observationally implied relation obtained by combining the \(F213\) lens flux with \(\theta_{\rm E}\), as computed from Equation~\eqref{eq:green_band} by propagating the distributions of \(m_{F213}\), \(\theta_{\rm E}\), \(D_{\rm S}\), and \(A_{F213}\) specified by the mock setup and recovery results. 
    The dark-green curve shows the mean relation, while the light-green band shows the corresponding 68\% credible interval. 
    Because the intrinsic \(M_{F213}\)--\(M_{\rm L}\) relation is tight in the M-dwarf regime, its intersection with \(M_{F213}^{\rm obs}(M_{\rm L})\) selects a narrow range of \(M_{\rm L}\) and \(M_{F213}\), yielding the vertical CMD coordinate of the lens in Figure~\ref{fig:cmd} and setting the absolute-magnitude slice used for the color--color dereddening and metallicity inference in Figure~\ref{fig:color_color_plot}. 
    }
    \label{fig:mass_luminosity_relation}
\end{figure}

%
In the Roman filter set, the relevant optical--NIR CMD is \(F062-F213\) versus \(M_{F213}\). 
Figure~\ref{fig:cmd} shows isochrones from the PAdova and TRieste Stellar Evolution Code (PARSEC) v1.2S \citep{Bressan2012,Chen2014} in this CMD, with color indicating metallicity and transparency indicating age. 
In the M-dwarf regime, where \mbox{\(M_{F213}\gtrsim5\)}~mag, isochrones of different metallicity separate strongly in \(F062-F213\), whereas isochrones of different age largely overlap. 
At brighter magnitudes, the CMD behaves differently. 
In the FGK-dwarf regime, isochrones of different metallicity are closely spaced, giving weaker metallicity sensitivity in optical--NIR color, while near the turnoff and subgiant branch both age and metallicity produce substantial CMD displacements. 

In the following subsections, we combine multi-band lens photometry with microlensing constraints to recover the intrinsic positions of M-dwarf lenses on this Roman optical--NIR CMD and thereby infer their metallicities. 
Because this work is primarily a feasibility demonstration of the method, we use the PARSEC isochrones as the CMD position--metallicity mapping throughout the principle illustration and mock-recovery experiments. 
The adopted PARSEC grid assumes a scaled-solar abundance pattern, so the metallicity label used throughout this work, \([{\rm M/H}]\), should be interpreted as the scaled-solar \([{\rm M/H}]\) required to reproduce the lens color--magnitude position.
Real applications will require an empirical calibration of the Roman-filter CMD position--metallicity mapping, using FGK+M binaries and/or stellar clusters, to account for theoretical-isochrone systematics and non-scaled-solar abundance patterns. 
We return to these chemical-interpretation and empirical-calibration issues in Section~\ref{subsec:4_1}. 


\subsection{Microlensing constraints on lens absolute magnitude}
\label{subsec:2_2}




The photometric metallicity method for M dwarfs has so far been applied mainly to nearby stars. 
To extend this method to microlensing lenses, we must recover two quantities that are usually straightforward for nearby stars but nontrivial for distant lenses. 
The first is the lens absolute magnitude, which determines the vertical CMD coordinate. 
The second is the dereddened optical--NIR color of the lens, which determines the horizontal CMD coordinate. 
This subsection focuses on the absolute-magnitude constraint, whereas the dereddening procedure is developed in Section~\ref{subsec:2_3}. 

A main difference from nearby-star applications is that microlensing lenses generally lack useful astrometric parallaxes. 
For nearby stars, distances can often be obtained from astrometric parallaxes, allowing apparent magnitudes to be converted into absolute magnitudes. 
Such astrometric parallax measurements are generally unavailable for microlensing lenses toward the Galactic bulge because of their large distances and faintness, but microlensing provides alternative mass--distance constraints. 
The relevant observables are the lens flux, the angular Einstein radius \(\theta_{\rm E}\), and the microlens-parallax vector \(\boldsymbol{\pi}_{\rm E}\). 
Each observable defines a relation between the lens mass \(M_{\rm L}\) and lens distance \(D_{\rm L}\), and combining any two of them can constrain \(M_{\rm L}\) and \(D_{\rm L}\), yielding the absolute-magnitude constraint needed for the vertical CMD coordinate. 

The first observable is the lens flux. 
As discussed in Section~\ref{subsec:2_1}, M-dwarf near-infrared absolute magnitudes depend only weakly on metallicity and even less on age, because the redward redistribution of flux partly compensates for the lower bolometric luminosity of metal-rich stars, and because main-sequence M dwarfs evolve negligibly over Galactic timescales. 
Thus a near-infrared absolute magnitude such as \(M_{F213}\) primarily traces \(M_{\rm L}\), as illustrated by the tight \(M_{F213}\)--\(M_{\rm L}\) relation from the PARSEC isochrones in the M-dwarf portion of Figure~\ref{fig:mass_luminosity_relation}, even across the broad metallicity and age range shown. 
For a measured lens apparent magnitude \(m_{F213}\), the lens-flux constraint is 
\begin{equation}
m_{F213}
=
M_{F213}(M_{\rm L}, [{\rm M/H}])
+
\mu(D_{\rm L})
+
A_{F213},
\label{eq:lens_flux_constraint}
\end{equation}
where \(\mu(D_{\rm L})=5\log_{10}(D_{\rm L}/10\,{\rm pc})\) is the lens distance modulus, and \(A_{F213}\) is the extinction to the lens in \(F213\). 
For a specified or internally inferred \(A_{F213}\), Equation~\eqref{eq:lens_flux_constraint} primarily defines a relation between \(M_{\rm L}\) and \(D_{\rm L}\), because the metallicity dependence of \(M_{F213}\) is weak for M dwarfs. 
The residual metallicity dependence of \(M_{F213}\), together with the internally inferred extinction \(A_{F213}\), is handled through the color--color dereddening and metallicity inference described in Section~\ref{subsec:2_3}. 

The second observable is the angular Einstein radius \(\theta_{\rm E}\), which sets the characteristic angular scale of the lensing event. 
It is related to the lens mass and lens--source relative parallax by
\begin{equation}
\theta_{\rm E}
=
\sqrt{\kappa M_{\rm L}\pi_{\rm rel}},
\quad
\pi_{\rm rel}
=
\pi_{\rm L}-\pi_{\rm S}
=
{\rm au}
\left(
\frac{1}{D_{\rm L}}-\frac{1}{D_{\rm S}}
\right),
\label{eq:thetaE_constraint}
\end{equation}
where \(D_{\rm S}\) is the source distance, \(\pi_{\rm rel}\) is the lens--source relative parallax, and \(\kappa=4G/(c^2{\rm au})\simeq 8.144~{\rm mas}~M_\odot^{-1}\). 
A measurement of \(\theta_{\rm E}\) therefore constrains \(M_{\rm L}\pi_{\rm rel}\), and hence defines another relation between \(M_{\rm L}\) and \(D_{\rm L}\), with \(D_{\rm S}\) supplied by the Galactic-model source-distance prior described in Section~\ref{subsec:3_1}.

For Roman planetary microlensing events, \(\theta_{\rm E}\) can be obtained through either of two primary routes. 
First, Roman multi-epoch high-resolution imaging can provide both the lens flux used in Equation~\eqref{eq:lens_flux_constraint} and the lens--source relative proper motion \(\mu_{\rm rel}\). 
Combined with the event timescale \(t_{\rm E}\) measured from the microlensing light curve, this gives \(\theta_{\rm E}=t_{\rm E}\mu_{\rm rel}\). 
Second, the light curve may provide a measurement of the normalized source radius \(\rho=\theta_\ast/\theta_{\rm E}\). 
With the source angular radius \(\theta_\ast\) inferred from the source color and magnitude using a color--surface brightness relation \citep[e.g.,][]{Adams2018}, this gives \(\theta_{\rm E}=\theta_\ast/\rho\). 
Roman simulations indicate that these routes should provide \(\theta_{\rm E}\) measurements for a substantial fraction of planetary microlensing events \citep{Penny2019,paper4_Terry2026}. 

The third observable is the microlens-parallax vector \citep{Gould1992,Gould2000,Gould2004}, 
\begin{equation}
    \boldsymbol{\pi}_{\rm E}=
    \frac{\pi_{\rm rel}}{\theta_{\rm E}}
    \frac{\boldsymbol{\mu}_{\rm rel}}{\mu_{\rm rel}}, 
\label{eq:piE_constraint}
\end{equation}
which is dimensionless. 
Although astrometric parallax of the lens star is generally unavailable, the same accelerated annual motion of the observer can make the apparent lens--source trajectory deviate from a straight line, sometimes leaving a detectable distortion in the microlensing light curve. 
The amplitude \(\pi_{\rm E}=\pi_{\rm rel}/\theta_{\rm E}\) is the angular lens--source position shift produced by a 1 au displacement of the observer, normalized by the angular Einstein radius. 
The direction of \(\boldsymbol{\pi}_{\rm E}\) is the direction of the lens--source relative proper motion, and therefore specifies how the observer's known annual displacement projects relative to the event trajectory. 
Together, the amplitude and direction determine the time-dependent perturbation to the lens--source relative-position vector in units of \(\theta_{\rm E}\), and hence the corresponding distortion of the microlensing light curve. 
When \(\boldsymbol{\pi}_{\rm E}\) is measured, its amplitude constrains \(\pi_{\rm rel}/M_{\rm L}\), thereby providing another relation between \(M_{\rm L}\) and \(D_{\rm L}\). 

Combining any two of these observables can constrain \(M_{\rm L}\) and \(D_{\rm L}\), and for Roman microlensing events the most frequently available pair is expected to be the lens flux and \(\theta_{\rm E}\). 
Figure~\ref{fig:mass_luminosity_relation} illustrates this case in the \(M_{F213}\)--\(M_{\rm L}\) plane. 
This combination produces an observationally implied \(M_{F213}\)--\(M_{\rm L}\) relation, \(M_{F213}^{\rm obs}(M_{\rm L})\), in which the \(\theta_{\rm E}\) constraint maps each trial \(M_{\rm L}\) to an associated \(D_{\rm L}\), and the lens-flux constraint then gives the implied \(F213\) absolute magnitude. 
Using the \(\theta_{\rm E}\) constraint of Equation~\eqref{eq:thetaE_constraint} to eliminate \(D_{\rm L}\) from the lens-flux constraint of Equation~\eqref{eq:lens_flux_constraint}, this relation is
\begin{equation}
    \begin{split}
    &M_{F213}^{\rm obs}(M_{\rm L})
    =\\
    &m_{F213}
    +
    5\log_{10}
    \left[
    10\,{\rm pc}
    \left(
    \frac{\theta_{\rm E}^2}{\kappa M_{\rm L}{\rm au}}
    +
    \frac{1}{D_{\rm S}}
    \right)
    \right]
    -
    A_{F213}.
    \end{split}
    \label{eq:green_band}
\end{equation}
The green band in Figure~\ref{fig:mass_luminosity_relation} shows this observationally implied relation for the fiducial bulge M-dwarf lens event used in the mock-recovery experiment in Section~\ref{sec:3_mock_recovery_experiments}. 
It is computed by propagating the distributions of \(m_{F213}\), \(\theta_{\rm E}\), \(D_{\rm S}\), and \(A_{F213}\) specified by the mock setup and recovery results, with the dark-green curve showing the mean relation and the light-green band showing the corresponding 68\% credible interval. 
In particular, the \(A_{F213}\) distribution is not adopted from an external extinction estimate, but is constrained by the internal color--color dereddening inference developed in Section~\ref{subsec:2_3}. 
Alongside this observationally implied relation, the PARSEC v1.2S isochrones shown in the same figure provide the intrinsic \(M_{F213}\)--\(M_{\rm L}\) relation set by stellar physics, spanning the same metallicity and age grid as in Figure~\ref{fig:cmd}. 
For the fiducial bulge M-dwarf lens event considered here, because the intrinsic \(M_{F213}\)--\(M_{\rm L}\) relation is tight in the M-dwarf regime, its intersection with \(M_{F213}^{\rm obs}(M_{\rm L})\) selects a narrow range of \(M_{\rm L}\) and \(M_{F213}\), yielding the vertical CMD coordinate needed to apply the M-dwarf photometric metallicity method. 
This lens-flux plus \(\theta_{\rm E}\) pair is expected to be the most frequently available because Roman high-resolution imaging should provide lens-flux measurements for many events, and because the routes discussed above should yield \(\theta_{\rm E}\) measurements for a substantial fraction of the relevant Roman microlensing events. 

The combination of lens-flux and \(\theta_{\rm E}\) constraints is strongest for distant bulge and inner-disk lenses and becomes weaker for relatively nearby lenses, but this distance dependence should not reduce the applicability of the method for most lenses, given the expected lens-distance distribution and the availability of \(\boldsymbol{\pi}_{\rm E}\) for nearby lenses. 
For distant bulge and inner-disk lenses, the \(M_{F213}^{\rm obs}(M_{\rm L})\) relation intersects the intrinsic \(M_{F213}\)--\(M_{\rm L}\) relation cleanly, as illustrated in Figure~\ref{fig:mass_luminosity_relation}. 
For relatively nearby lenses, however, \(M_{F213}^{\rm obs}(M_{\rm L})\) can become less orthogonal to the intrinsic relation, leaving a broader allowed range of \(M_{\rm L}\), \(D_{\rm L}\), and \(M_{F213}\). 
Nevertheless, this limitation should affect only a minority of the expected sample, because simulations predict that the majority of Roman lenses will lie in the Galactic bulge \citep{Penny2019,paper4_Terry2026}. 
Moreover, the same nearby lenses for which the lens-flux plus \(\theta_{\rm E}\) constraints provide weaker leverage have larger \(\pi_{\rm rel}\), and hence larger \(\pi_{\rm E}\) at fixed \(M_{\rm L}\), making microlens-parallax measurements generally more accessible than for distant lenses. 
When \(\boldsymbol{\pi}_{\rm E}\) is available, it provides an additional, largely orthogonal mass--distance constraint that complements the lens-flux and \(\theta_{\rm E}\) constraints, yielding a cleaner determination of \(M_{\rm L}\), \(D_{\rm L}\), and the vertical CMD coordinate \(M_{F213}\). 

\begin{figure*}
    \centering
    \includegraphics[width=0.8\textwidth]{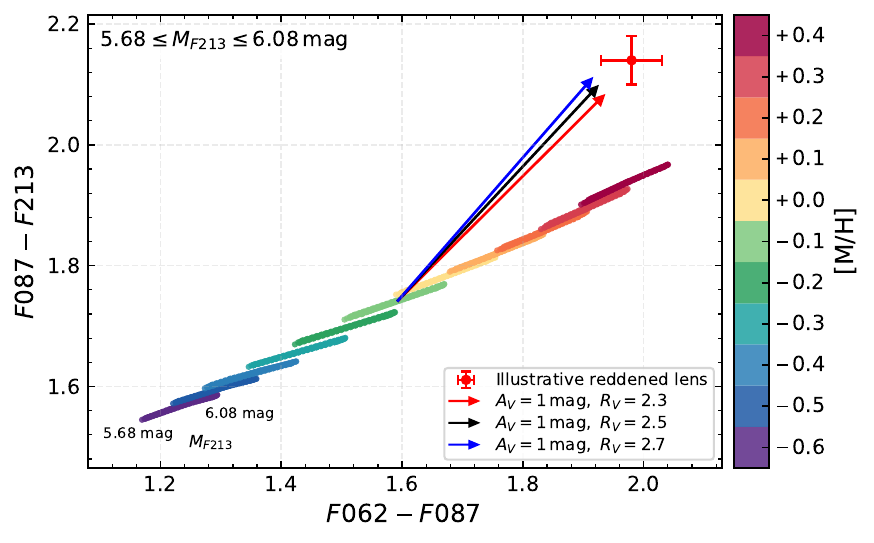}
    \caption{
    Color--color dereddening and metallicity inference in the Roman \((F062-F087,\,F087-F213)\) color--color diagram. 
    The red point with error bars shows an illustrative observed position of the reddened lens and the expected photometric uncertainties. The colored segments show the intrinsic PARSEC v1.2S stellar locus, color-coded by metallicity. 
    Only the portion of the locus with \(5.68<M_{F213}<6.08\) mag is shown, corresponding to the 68\% credible interval in \(M_{F213}\) from the mock recovery of the fiducial event shown in Figure~\ref{fig:mass_luminosity_relation}. The lens-flux plus \(\theta_{\rm E}\) constraints in that figure illustrate the origin of this narrow absolute-magnitude range, which itself is not determined independently of extinction. The restricted locus shown here therefore provides a graphical illustration of the inference; in the full mock recovery, \(M_{F213}\) and the lens extinction are inferred jointly.
    Along each segment, the left and right endpoints correspond to \(M_{F213}=5.68\) and \(6.08\) mag, respectively, as labeled for the \([{\rm M/H}]=-0.6\) example. 
    The arrows show representative inner-Galaxy reddening-vector directions computed for \(R_V=2.3,2.5,\) and \(2.7\). 
    The arrows are normalized to \(A_V=1\)~mag only to indicate direction, whereas Roman microlensing fields have mean extinctions to the bulge of order \(A_V\sim4\) mag.  
    Moving the observed lens position backward along the reddening vector until it intersects the intrinsic locus simultaneously constrains the lens extinction and metallicity. 
    }
\label{fig:color_color_plot}
\end{figure*}

\begin{figure}
    \centering
    \includegraphics[width=\columnwidth]{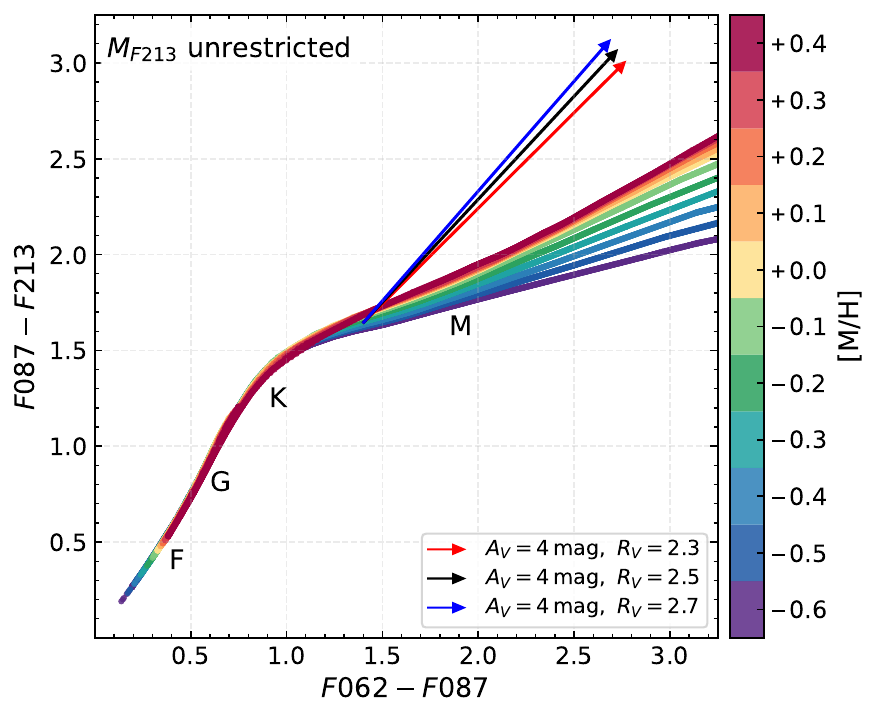}
    \caption{
    Roman \((F062-F087,\,F087-F213)\) color--color diagram without imposing the \(M_{F213}\) restriction from the microlensing constraints. 
    The colored curves show the intrinsic PARSEC v1.2S dwarf locus, color-coded by metallicity, with approximate F, G, K, and M regimes labeled. 
    The same reddening-vector directions as in Figure~\ref{fig:color_color_plot} are shown, here normalized to \(A_V=4\) mag. 
    The figure illustrates two points. 
    First, the FGK-dwarf portion of the locus is nearly aligned with the reddening vectors, whereas the M-dwarf portion forms a larger angle with them. 
    Second, without the \(M_{F213}\) restriction, each metallicity sequence spans a broad range in color--color space, so loci of different metallicities can overlap and dereddening can allow multiple \(M_{F213}\)--\([{\rm M/H}]\) solutions rather than a unique metallicity. 
    Restricting \(M_{F213}\) with the microlensing constraints selects only a short slice of this locus, enabling a clear metallicity inference from the dereddening intersection, as shown in Figure~\ref{fig:color_color_plot}. 
    }
    \label{fig:color_color_no_MF213_constrain}
\end{figure}

\subsection{Color--color dereddening and metallicity inference}
\label{subsec:2_3}

The microlensing constraints provide the main handle on the vertical CMD coordinate, but the observed optical--NIR color must still be dereddened to recover the intrinsic lens color, i.e., the horizontal CMD coordinate. 
Photometric metallicity methods for M dwarfs are typically applied to stars within tens to a few hundreds of parsecs, where the extinction is generally low and not a dominant term in the metallicity error budget. 
Microlensing lenses toward the Galactic bulge are different, since they are often several kiloparsecs away and lie behind a substantial fraction of the Galactic-plane dust column. 
Representative Roman microlensing fields have mean optical extinctions to the bulge of order \(A_V\sim4\) mag, with a central 95\% range across sightlines of roughly \(A_V\sim2\mbox{--}7\) mag and significant angular and distance-dependent structure \citep{dustmaps,Zucker2025}. 
Thus, a useful photometric microlensing-lens metallicity method must determine the extinction to the lens accurately enough to recover the dereddened lens color.

Assigning the lens extinction from external dust information is inadequate for the accurate optical--NIR dereddening required here. 
A common approach for estimating the foreground-lens extinction in microlensing is to scale the bulge red-clump extinction measured at the event sightline by an assumed line-of-sight dust distribution, often an exponential model \citep[e.g.,][]{HST_Bennett2015_OGLE-2005-BLG-169, Batista2015_ob05169}. 
Such a smooth line-of-sight prescription is a simplified approximation, because the dust distribution toward the inner Galaxy is highly structured in three dimensions \citep{Vergely2022_3Ddustmap,Zucker2025}. 
Nevertheless, this approximation can be adequate for many near-infrared lens-flux analyses, where near-infrared extinction corrections are relatively small, so that their uncertainties often remain subdominant in the lens-flux error budget. 
However, it is insufficient for accurately dereddening optical--NIR colors, where the optical extinction is much larger and directly affects the inferred metallicity.
Modern three-dimensional dust maps now cover the relevant bulge microlensing fields and extend to \(\sim10~\mathrm{kpc}\) \citep[e.g.,][]{Zucker2025}, providing a much improved description of the distance-dependent dust distribution. 
Even so, their limited angular resolution, distance resolution, and systematic uncertainties still make them insufficient for assigning a precise extinction to an individual microlensing lens. 

Instead of deriving the lens extinction from external dust information, we infer it from the lens's own multi-band photometry together with the microlensing constraints. 
The extinction to the lens is therefore treated as a fitted quantity, rather than a value read from a dust model or dust map. 
As shown in Figure~\ref{fig:color_color_plot}, this requires three-band lens photometry, here Roman $F062$, $F087$, and $F213$, to place the observed lens in the \((F062-F087,\,F087-F213)\) color--color diagram.
An illustrative observed position of a reddened lens and the expected photometric uncertainty are shown to demonstrate the dereddening procedure. 
The same figure also shows the unreddened stellar locus from PARSEC v1.2S \citep{Bressan2012,Chen2014}, color-coded by metallicity. 
Because the lens-flux plus \(\theta_{\rm E}\) constraints described in Section~\ref{subsec:2_2} restrict \(M_{F213}\) to a narrow interval, the locus shown in Figure~\ref{fig:color_color_plot} is limited to the corresponding absolute-magnitude slice from Figure~\ref{fig:mass_luminosity_relation}. 
With the observed color--color position and the allowed intrinsic locus specified, the remaining ingredient is which direction extinction displaces the lens in this diagram. 
Because dust absorption and scattering are wavelength dependent, the reddening of different colors is correlated. 
For a given extinction curve, increasing extinction therefore displaces a star along a specific direction in color--color space, which defines the reddening-vector direction. 
The arrows in Figure~\ref{fig:color_color_plot} show representative inner-Galaxy reddening-vector directions. Their calculation and calibration are discussed in Section~\ref{subsec:2_4}. 

The dereddening is carried out directly in the color--color diagram. 
We deredden the lens by moving its observed position backward along the reddening vector until it intersects the intrinsic M-dwarf locus, so that the displacement along the vector gives the lens extinction, while the intersection gives the dereddened color and hence the lens metallicity. 
This construction is enabled by the fact that the M-dwarf locus in optical--NIR color--color space is not parallel to typical reddening vectors \citep[see, e.g., Section~7.5 of][]{LSST_science_book}, in contrast to the nearly aligned FGK locus \citep{Covey2007,Sale2009,Berry2012}, as shown in Figure~\ref{fig:color_color_plot} and \ref{fig:color_color_no_MF213_constrain}. 

The microlensing constraints are essential for turning this color--color dereddening into a metallicity measurement. 
Figure~\ref{fig:color_color_no_MF213_constrain} shows the same Roman color--color diagram without imposing the \(M_{F213}\) restriction. 
Without microlensing constraints, \(M_{F213}\) remains free to vary along the M-dwarf sequence, so a change in metallicity can be partly compensated by a change in \(M_{F213}\), allowing multiple \((M_{F213},[{\rm M/H}])\) pairs to land near the same position in the color--color diagram and preventing a unique metallicity inference. 
The measured \(F213\) lens flux, combined with the available microlensing constraints, whether \(\theta_{\rm E}\), \(\boldsymbol{\pi}_{\rm E}\), or both, restricts the allowed \(M_{F213}\) interval. 
Within the short \(M_{F213}\) slice shown in Figure~\ref{fig:color_color_plot}, the remaining variation of the intrinsic M-dwarf locus is then dominated by metallicity, making the lens metallicity recoverable from the dereddened colors. 

The color--color dereddening and metallicity inference described above can be summarized as a joint fit to the lens photometry and microlensing constraints. 
Schematically, for the three Roman bands we solve
\begin{equation}
\boldsymbol{m}
=
\boldsymbol{M}\!\left(M_{\rm L},[{\rm M/H}]\right)
+
\mu(D_{\rm L})
+
\boldsymbol{A}(A_V,R_V),
\end{equation}
together with the available microlensing constraint, either the \(\theta_{\rm E}\) constraint from Equation~\eqref{eq:thetaE_constraint}, or the \(\boldsymbol{\pi}_{\rm E}\) constraint from Equation~\eqref{eq:piE_constraint}, or both. 
Here \(\boldsymbol{m}=(m_{F062},m_{F087},m_{F213})\) is the observed lens-magnitude vector, \(\boldsymbol{M}\) is the absolute-magnitude vector in the same bands, \(\mu(D_{\rm L})\) is the lens distance modulus added to each band, and \(\boldsymbol{A}\) is the extinction vector computed from Equation~\eqref{eq:bandpass_extinction}. 

We refer to this joint use of multi-band lens photometry and microlensing mass--distance constraints to infer the lens metallicity as photometric--microlensing metallicity inference. 
This is related in spirit to photo-astrometric stellar-parameter inference and dust-mapping methods \citep[e.g.,][]{Bailer-Jones2011,Green2019,Anders2022}, but applied in the M-dwarf regime where the molecular-opacity-driven optical--NIR colors together with the non-parallel orientation of the M-dwarf locus relative to the reddening vector provide metallicity leverage, with microlensing observables replacing direct astrometric parallax as the geometric constraint.

\subsection{Reddening-vector direction}
\label{subsec:2_4}

The inference described above treats the extinction to the lens as a fitted quantity, but it still requires an independently specified reddening-vector direction in the adopted color--color diagram. 
In the Roman \((F062-F087,\,F087-F213)\) plane, this direction can be written as 
\begin{equation}
    S_{\rm reddening}
    \equiv
    \frac{E(F087-F213)}{E(F062-F087)}
    =
    \frac{A_{F087}-A_{F213}}{A_{F062}-A_{F087}} .
    \label{eq:reddening_vector_slope}
\end{equation}
The fitted extinction determines how far the lens is displaced along the reddening vector, while \(S_{\rm reddening}\) sets the reddening-vector direction and therefore controls the angle between the reddening vector and the M-dwarf locus, affecting how cleanly extinction and metallicity can be separated. 

The value of \(S_{\rm reddening}\) is determined by the wavelength dependence of extinction, and can therefore be computed from an extinction curve after integrating over the relevant bandpasses. 
In this paper, we adopt the \citet{CCM89} extinction-curve family (hereafter \citetalias{CCM89}) to generate representative reddening-vector directions, while the empirical Roman-band calibration needed for real applications is discussed below. 
The \citetalias{CCM89} family is parameterized by \(R_V\equiv A_V/E(B-V)\), where smaller \(R_V\) corresponds to a steeper optical extinction curve. 
Although \(R_V=3.1\) is commonly adopted for the diffuse interstellar medium \citep{CCM89}, it is not representative of many inner-Galaxy sightlines. 
Toward the Galactic bulge, \citet{Nataf2013} found that the optical--NIR reddening law is better described by \(R_V\simeq 2.5\) within the \citetalias{CCM89} family, with a dispersion of order \(\sigma_{R_V}\simeq0.2\). 
We therefore use \(R_V=2.3,2.5,\) and \(2.7\) to illustrate representative inner-Galaxy reddening-vector directions in Figure~\ref{fig:color_color_plot} and \ref{fig:color_color_no_MF213_constrain}. 

For each value of \(R_V\), we compute the bandpass-integrated extinction in each Roman filter.  
For a bandpass \(X\), the extinction is
\begin{equation}
\begin{split}
    &A_X(A_V,R_V)
    = \\
    &-2.5\log_{10}
    \left[
    \frac{
    \int \lambda F_\lambda(\lambda)\,
    10^{-0.4A(\lambda;A_V,R_V)}
    T_X(\lambda)\,d\lambda
    }{
    \int \lambda F_\lambda(\lambda)\,
    T_X(\lambda)\,d\lambda
    }
    \right],
\end{split}
\label{eq:bandpass_extinction}
\end{equation}
where \(F_\lambda(\lambda)\) is the stellar spectral flux density, \(T_X(\lambda)\) is the Roman WFI throughput curve for filter \(X\), and \(A(\lambda;A_V,R_V)\) is the wavelength-dependent extinction for a given \(A_V\) and \(R_V\), with the factor of \(\lambda\) accounting for the photon-counting response of the detector. 
We evaluate Equation~\eqref{eq:bandpass_extinction} using PHOENIX M-dwarf spectra \citep{PHOENIX}, the \href{https://github.com/RomanSpaceTelescope/roman-technical-information/tree/v1.3/data/WideFieldInstrument/Imaging/EffectiveAreas}{Roman WFI imaging effective-area curves}, \texttt{synphot} \citep{synphot}, and \texttt{dust\_extinction} \citep{dust_extinction}. 
The resulting normalized extinctions, \(A_X/A_V\), give the relative reddening in the two colors entering Equation~\eqref{eq:reddening_vector_slope}, and therefore determine \(S_{\rm reddening}\).

The resulting inner-Galaxy reddening-vector directions shown in Figures~\ref{fig:color_color_plot} and \ref{fig:color_color_no_MF213_constrain} demonstrate that the vectors are not parallel to the M-dwarf locus, which is the key property enabling the dereddening procedure described in Section~\ref{subsec:2_3}. 
The same figures also illustrate that the reddening-vector direction has a plausible range, so uncertainty in the direction can contribute to the metallicity error budget. 
However, the range shown here reflects variations measured over a large inner-Galaxy region, whereas each microlensing event has a precisely known sightline. 
In applying the method to Roman data, one can use a locally calibrated reddening-vector direction for that sightline, \(S_{\rm reddening}(l,b)\), or for that sightline and distance, \(S_{\rm reddening}(l,b,D)\), as described below, which should reduce the reddening-vector direction uncertainty. 

The \citetalias{CCM89}-based calculation is useful for illustrating the method, but it should not be regarded as the reddening-vector direction ultimately adopted for Roman data. 
Indeed, \citet{Nataf2016} showed that the observed inner-Galaxy extinction-curve variations cannot be fully reproduced by commonly used extinction-curve families, regardless of how \(R_V\) is varied. 
Thus, the Roman-band \(S_{\rm reddening}\) should be empirically calibrated from Roman data themselves, rather than inferred through an analytic \(R_V\)-based prescription.

A two-dimensional empirical calibration for the GBTDS fields, \(S_{\rm reddening}(l,b)\), can be obtained from the bulge red clump. 
Red-clump stars are widely used as standard crayons because of their relatively consistent intrinsic colors \citep{Gonzalez2012,Nataf2013,Surot2020}, so their displacement in the Roman color--color diagram can be used to measure the reddening-vector direction along each sightline. 
This calibration should be adequate for Roman microlensing lenses lying close to the bulge distance, which should share a similar reddening-vector direction with the bulge red clump along the same sightline. 

The reddening-vector direction can also depend on distance, so for lenses lying well in front of the bulge red clump, a three-dimensional empirical calibration, \(S_{\rm reddening}(l,b,D)\), is more appropriate. 
Such a distance-resolved calibration for GBTDS fields can be constructed by combining planned Roman grism spectroscopic snapshots \citep{ROTAC_final_report} with multi-band Roman imaging, in a similar spirit to the Gaia-XP-based three-dimensional reddening-law mapping of \citet{ZhangGreen2025}, with the spectra helping to separate stellar type from reddening-law variation. 
This would supply the \(S_{\rm reddening}\) needed for microlensing-lens metallicity inference and provide a Roman-band diagnostic of spatial variations in reddening law toward the Galactic bulge.



\section{Mock-recovery Experiments}\label{sec:3_mock_recovery_experiments}
In this section, we assess the expected performance of the photometric--microlensing metallicity method through a mock-recovery experiment. 
We first present the \texttt{MetalLens} inference framework, which implements the method, then construct a fiducial mock Roman microlensing event and examine the recovery of its physical parameters. 

\subsection{The \texttt{MetalLens} inference framework}
\label{subsec:3_1}

We develop \texttt{MetalLens}\footnote{\url{https://github.com/zhangjiyuan22/MetalLens}}, an inference code that implements the photometric--microlensing metallicity method described in Section~\ref{sec:2_principle}.
For a measured observable vector \(\hat{\boldsymbol{d}}\), which consists of multi-band lens photometry and microlensing observables, the posterior distribution of the physical parameters of the event is
\begin{equation}
p(\boldsymbol{\Theta}|\hat{\boldsymbol{d}})
\propto
p(\hat{\boldsymbol{d}}|\boldsymbol{\Theta})p(\boldsymbol{\Theta}),
\end{equation}
where \(\boldsymbol{\Theta}\) denotes the physical parameters of the event. 
We approximate this posterior by importance sampling. 
Operationally, \texttt{MetalLens} draws a large number (e.g., \(\mathcal{O}(10^9)\)) of simulated events \(\boldsymbol{\Theta}_i\) from the prior \(p(\boldsymbol{\Theta})\), forward-models the observables \(\boldsymbol{d}(\boldsymbol{\Theta}_i)\) for each event \(i\), and assigns each event a weight \(w_i\) proportional to \(p(\hat{\boldsymbol{d}}|\boldsymbol{d}(\boldsymbol{\Theta}_i))\). 
The posterior distribution is then represented by the weighted sample \(\{\boldsymbol{\Theta}_i,w_i\}\). 
We verify convergence of this posterior estimate using importance-sampling diagnostics, including the effective sample size and the fraction of total weight contributed by the highest-weight events. 

For each simulated event \(i\), the parameter vector is 
\begin{equation}
\boldsymbol{\Theta}_i
\equiv
\left(
M_{\rm L},D_{\rm L},[{\rm M/H}],\tau,A_V,
\boldsymbol{v}_{{\rm L},\perp},\boldsymbol{v}_{{\rm S},\perp},
D_{\rm S},R_V
\right),
\end{equation}
where \(\tau\) is the lens age, \(\boldsymbol{v}_{{\rm L},\perp}\) and \(\boldsymbol{v}_{{\rm S},\perp}\) are the transverse velocities of the lens and source, and the remaining notation follows Section~\ref{sec:2_principle}. 
As discussed in Section~\ref{subsec:2_4}, real applications should use an empirically calibrated Roman-band reddening-vector direction rather than an analytic \(R_V\)-based prescription. 
For the mock-recovery experiments, however, we retain the analytic \(R_V\)-based prescription to specify representative inner-Galaxy reddening-vector directions and to propagate uncertainty in the reddening-vector direction into the inferred metallicity error budget. 
%

Except for \(D_{\rm S}\) and \(R_V\), all parameters are drawn from broad uniform priors rather than informative Galactic population priors. 
This choice is deliberate for two reasons. 
First, photometric and microlensing observables provide direct constraints on \(M_{\rm L}\), \(D_{\rm L}\), \([{\rm M/H}]\), and \(A_V\) on an event-by-event basis. 
Second, imposing informative Galactic population priors on these quantities would bias the population-level lens-parameter statistics that we aim to measure. 
The remaining uniform-prior parameters enter as nuisance variables, because \(\tau\) has little effect on the M-dwarf CMD position over the main sequence, while the individual transverse velocities enter the likelihood only through the relative proper motion, which is directly constrained by the observations.\footnote{In the actual implementation, \texttt{MetalLens} does not explicitly draw \(\boldsymbol{v}_{{\rm L},\perp}\) and \(\boldsymbol{v}_{{\rm S},\perp}\). 
Because these velocities enter the likelihood only through \(\boldsymbol{\mu}_{\rm rel,H}\), we instead draw \(\boldsymbol{\mu}_{\rm rel,H}\) for each simulated event directly from the two-dimensional Gaussian distribution defined by its value and covariance measured from high-resolution imaging, and use it to compute \(\boldsymbol{\mu}_{\rm rel,Roman}\) and the corresponding \(t_{\rm E}\). 
The proper-motion term is then omitted from Equation~\eqref{eq:weight}, so that the same proper-motion measurement does not contribute to the weight twice. 
This is equivalent to drawing individual transverse velocities and weighting the simulated event by the observed \(\boldsymbol{\mu}_{\rm rel,H}\), but is much more efficient. }

The only externally informative priors are those assigned to \(D_{\rm S}\) and to the reddening-vector direction, parameterized here by \(R_V\). 
The lens photometry and microlensing observables do not by themselves determine these quantities. 
Moreover, they must be specified externally to translate the observables into constraints on the target parameters \(M_{\rm L}\), \(D_{\rm L}\), \([{\rm M/H}]\), and \(A_V\). 
Specifically, for a given \(D_{\rm L}\), \(D_{\rm S}\) is needed to determine \(\pi_{\rm rel}\), which enters the microlensing mass--distance constraints, while the reddening-vector direction specifies the direction along which the observed lens colors are dereddened in the color--color plane. 
This use of externally informative priors is justified because \(D_{\rm S}\) and the reddening-vector direction are not part of the target statistics. 
In practice, we marginalize over their prior distributions, so that uncertainty in \(D_{\rm S}\) and the reddening-vector direction is propagated into the inferred uncertainties of \(M_{\rm L}\), \(D_{\rm L}\), \([{\rm M/H}]\), and \(A_V\). 
For \(D_{\rm S}\), we adopt a line-of-sight-dependent prior based on the event-rate-weighted stellar number density,
\begin{equation}
    p(D_{\rm S}|l,b)
    \propto
    n_\ast(D_{\rm S})
    \int_0^{D_{\rm S}} n_\ast(D)\,dD,
    \label{eq:DS_prior}
\end{equation}
where \(n_\ast(D)\) is the stellar number density at distance \(D\) along the event sightline \((l,b)\), taken from the Galactic model of \citet{Yang2021_GalacticModel}. 
For the central sightline of the five contiguous GBTDS fields, this yields a mean and standard deviation of \(D_{\rm S}=9.0\pm1.3~\mathrm{kpc}\), with slight variations across the field. 
For the reddening-vector direction, motivated by the inner-Galaxy reddening law of \citet{Nataf2013}, we adopt a Gaussian prior centered at \(R_V=2.5\), but with a narrower width than the dispersion of \(\sigma_{R_V}\simeq0.2\) reported by \citet{Nataf2013}, because that dispersion reflects variations over extended inner-Galaxy fields, whereas each microlensing event has a precisely known sightline and real applications can use an empirical calibration of the reddening-vector direction local to that sightline, as discussed in Section~\ref{subsec:2_4}. 
We therefore adopt a standard deviation of \(\sigma_{R_V}=0.1\) as a proxy for the residual uncertainty in the locally calibrated reddening-vector direction. 

For each simulated event \(i\), \texttt{MetalLens} then forward-models the predicted photometric and microlensing observables \(\boldsymbol{d}(\boldsymbol{\Theta}_i)\), which is later compared with the measured observable vector \(\hat{\boldsymbol{d}}\) to determine its weight \(w_i\). 
Here we consider a common case for \(\hat{\boldsymbol{d}}\), in which the three-band lens fluxes are measured, the \(\theta_{\rm E}\) constraint is supplied by \(t_{\rm E}\) and the lens--source relative proper motion, and no \(\boldsymbol{\pi}_{\rm E}\) measurement is assumed. 
The lens fluxes and the heliocentric lens--source relative proper motion \(\boldsymbol{\mu}_{\rm rel,H}\) are measured from high-resolution imaging, while \(t_{\rm E}\) is obtained from fitting the microlensing light curve. 
The corresponding predicted observable vector \(\boldsymbol{d}(\boldsymbol{\Theta}_i)\), with the same components as \(\hat{\boldsymbol{d}}\), is therefore 
\begin{equation}
\boldsymbol{d}(\boldsymbol{\Theta}_i)
=
\left(
c_{F062-F087},
c_{F087-F213},
m_{F213},
\boldsymbol{\mu}_{\rm rel,H},
t_{\rm E}
\right),
\label{eq:observable_vector}
\end{equation}
where \(c_{F062-F087}\) and \(c_{F087-F213}\) are lens colors, defined as \(c_{F062-F087}\equiv m_{F062}-m_{F087}\) and \(c_{F087-F213}\equiv m_{F087}-m_{F213}\). 
The photometric part of this vector is written as the triplet 
\((c_{F062-F087},c_{F087-F213},m_{F213})\), which is an equivalent reparameterization of the three-band lens photometry. 
This form is chosen because the two color components specify the lens position in the color--color plane, where the metallicity and extinction information are most directly represented.

To predict the \(\boldsymbol{\mu}_{\rm rel,H}\) and \(t_{\rm E}\) components of \(\boldsymbol{d}(\boldsymbol{\Theta}_i)\), \texttt{MetalLens} first computes \(\pi_{\rm rel}\) and \(\theta_{\rm E}\) for each simulated event from Equation~\eqref{eq:thetaE_constraint}. 
The heliocentric relative proper motion \(\boldsymbol{\mu}_{\rm rel,H}\) is related to the lens and source transverse velocities by
\begin{equation}
\boldsymbol{\mu}_{\rm rel,H}
=
\frac{\boldsymbol{v}_{{\rm L},\perp}-\boldsymbol{v}_{\odot,\perp}}{D_{\rm L}}
-
\frac{\boldsymbol{v}_{{\rm S},\perp}-\boldsymbol{v}_{\odot,\perp}}{D_{\rm S}},
\label{eq:murel_H_forward}
\end{equation}
where \(\boldsymbol{v}_{\odot,\perp}\) is the solar transverse velocity. 
This heliocentric relative proper motion is then converted to the Roman-centric frame,
\begin{equation}
\boldsymbol{\mu}_{\rm rel,Roman}
=
\boldsymbol{\mu}_{\rm rel,H}
-
\frac{\boldsymbol{v}_{{\rm Roman},\perp}\pi_{\rm rel}}{{\rm au}},
\label{eq:murel_Roman}
\end{equation}
where \(\boldsymbol{v}_{{\rm Roman},\perp}\) is Roman's projected velocity relative to the Sun at the time of peak magnification. 
The predicted event timescale is then
\begin{equation}
t_{\rm E}
=
\frac{\theta_{\rm E}}{\mu_{\rm rel,Roman}}. 
\label{eq:tE_forward}
\end{equation}
The observable set considered here does not include a measured \(\boldsymbol{\pi}_{\rm E}\). 
If such a measurement is available, \texttt{MetalLens} can also predict \(\boldsymbol{\pi}_{\rm E}\) from Equation~\eqref{eq:piE_constraint} and add the corresponding likelihood term to Equation~\eqref{eq:weight}. 

\texttt{MetalLens} then predicts the photometric components of \(\boldsymbol{d}(\boldsymbol{\Theta}_i)\). 
As discussed in Section~\ref{subsec:2_1} and later in Section~\ref{subsec:4_1}, real applications will require an empirically calibrated Roman-filter CMD position--metallicity mapping to account for theoretical-isochrone systematics and abundance-pattern effects such as \(\alpha\)-enhancement at low \([{\rm Fe/H}]\). 
For the mock-recovery experiments, however, we instead use the PARSEC isochrones as a feasibility demonstration of the method, which provide a theoretical mapping from \((M_{\rm L},[{\rm M/H}],\tau)\) to the Roman-band absolute magnitudes. 
In real applications, this theoretical mapping can be replaced by an empirical CMD position--metallicity mapping, together with a near-infrared mass--luminosity relation relating \(M_{F213}\) to \(M_{\rm L}\); the specific choice of the latter, as well as intrinsic scatter around it, should have limited influence on the inferred \(M_{F213}\), and hence on the inferred metallicity, because over the photometrically accessible M-dwarf mass range targeted by this method, the observationally implied \(M_{F213}^{\rm obs}(M_{\rm L})\) relation in Figure~\ref{fig:mass_luminosity_relation} primarily confines the allowed \(M_{F213}\) range. 
For each sampled \((M_{\rm L},[{\rm M/H}],\tau)\), \texttt{MetalLens} interpolates the PARSEC grid to obtain the lens absolute magnitudes \(M_{F062}\), \(M_{F087}\), and \(M_{F213}\). 
For the sampled \((A_V,R_V)\), it computes the bandpass-integrated extinctions \(A_{F062}\), \(A_{F087}\), and \(A_{F213}\) using Equation~\eqref{eq:bandpass_extinction}. 
The predicted photometric observables are then
\begin{equation}
c_{F062-F087}=(M_{F062}-M_{F087})+(A_{F062}-A_{F087}),
\label{eq:color_F062_F087}
\end{equation}
\begin{equation}
c_{F087-F213}=(M_{F087}-M_{F213})+(A_{F087}-A_{F213}),
\label{eq:eq:color_F087_F213}
\end{equation}
\begin{equation}
m_{F213}=M_{F213}+\mu(D_{\rm L})+A_{F213}, 
\label{eq:mF213_forward}
\end{equation}
where \(\mu(D_{\rm L})\) is the lens distance modulus. 

For each simulated event \(i\), \texttt{MetalLens} then assigns a weight \(w_i\) proportional to \(p(\hat{\boldsymbol{d}}|\boldsymbol{d}(\boldsymbol{\Theta}_i))\). 
For a measured observable vector \(\hat{\boldsymbol{d}}\) with Gaussian uncertainties described by the covariance matrix \(\boldsymbol{C}_{\hat{\boldsymbol{d}}}\), this likelihood is
\begin{equation}
w_i
\propto
p(\hat{\boldsymbol{d}}|\boldsymbol{d}(\boldsymbol{\Theta}_i))
=
{\cal N}(
\hat{\boldsymbol{d}}
\mid
\boldsymbol{d}(\boldsymbol{\Theta}_i),
\boldsymbol{C}_{\hat{\boldsymbol{d}}}),
\label{eq:full_gaussian_likelihood}
\end{equation}
where \({\cal N}(\boldsymbol{x}|\boldsymbol{\mu},\boldsymbol{C})\) denotes the multivariate Gaussian probability density with mean \(\boldsymbol{\mu}\) and covariance matrix \(\boldsymbol{C}\), evaluated at \(\boldsymbol{x}\). 
For the observable set in Equation~\eqref{eq:observable_vector}, adopting a diagonal approximation to \(\boldsymbol{C}_{\hat{\boldsymbol{d}}}\) gives 
\begin{align}
w_i \propto &\,
{\cal N}_1\left(\hat{c}_{F062-F087}\mid c_{F062-F087,i},\sigma_{F062-F087,\mathrm{tot}}\right)\nonumber\\
&\times
{\cal N}_1\left(\hat{c}_{F087-F213}\mid c_{F087-F213,i},\sigma_{F087-F213,\mathrm{tot}}\right)\nonumber\\
&\times
{\cal N}_1\left(\hat{m}_{F213}\mid m_{F213,i},\sigma_{m_{F213}}\right)\nonumber\\
&\times
{\cal N}_2\left(\hat{\boldsymbol{\mu}}_{\rm rel,H}\mid \boldsymbol{\mu}_{{\rm rel,H},i},\boldsymbol{C}_{\boldsymbol{\mu}_{\rm rel,H}}\right)\nonumber\\
&\times
{\cal N}_1\left(\hat{t}_{\rm E}\mid t_{{\rm E},i},\sigma_{t_{\rm E}}\right),
\label{eq:weight}
\end{align}
where \({\cal N}_1\) and \({\cal N}_2\) denote one- and two-dimensional Gaussian probability densities, respectively. 
Here, the total color uncertainties entering Equation~\eqref{eq:weight} are written as quadrature sums of a photometric-uncertainty component and an intrinsic-scatter component, 
\begin{equation}
\sigma_{F062-F087, \rm{tot}}^2
=
\sigma_{F062-F087, \rm{obs}}^2
+
\sigma_{F062-F087, \rm{int}}^2,
\label{eq:sigma_total_1}
\end{equation}
\begin{equation}
\sigma_{F087-F213, \rm{tot}}^2
=
\sigma_{F087-F213, \rm{obs}}^2
+
\sigma_{F087-F213, \rm{int}}^2. 
\label{eq:sigma_total_2}
\end{equation}
The \(\sigma_{\rm obs}\) terms denote the photometric uncertainties in the two colors, while the \(\sigma_{\rm int}\) terms describe the intrinsic astrophysical scatter around the empirical CMD position--metallicity relation expected for real applications. 
The latter accounts for the fact that even at the same CMD position, stars do not necessarily have the same metallicity, for example because of rotation, magnetic activity, abundance-pattern variations, or other stellar effects \citep{Neves2012, Duque-Arribas2023, Rains2024}. 
Including \(\sigma_{\rm int}\) therefore propagates this intrinsic scatter into the recovered metallicity uncertainty. 
Separately, in real applications, the multi-band lens fluxes and the \(\boldsymbol{\mu}_{\rm rel,H}\) common to all bands will be inferred jointly from the multi-band high-resolution images, and the resulting full covariance matrix of these quantities should be used in the weighting rather than the diagonal approximation in Equation~\eqref{eq:weight}. 

\subsection{A fiducial mock event}
\label{subsec:3_2}

In this subsection, we construct a fiducial mock Roman microlensing event by specifying its physical parameters, deriving the corresponding observables, and assigning realistic measurement uncertainties. 
These observables are then used as input to \texttt{MetalLens}, and the recovery of the physical parameters is presented in Section~\ref{subsec:3_3}. 

We adopt a bulge M-dwarf lens for the fiducial event, because simulations predict that the majority of Roman lenses will lie in the Galactic bulge \citep{Penny2019,paper4_Terry2026} and the method developed here targets M-dwarf lenses. 
Specifically, we adopt a lens with \(M_{\rm L}=0.5\,M_\odot\), \(D_{\rm L}=7.5~\mathrm{kpc}\), \([{\rm M/H}]=+0.1\), \(\tau=8~\mathrm{Gyr}\), and \(A_V=4.0~\mathrm{mag}\). 
The adopted lens distance is approximately the peak of the Roman lens-distance distribution predicted by \citet{paper4_Terry2026}, while the adopted extinction is roughly the mean cumulative extinction out to \(7.5~\mathrm{kpc}\) across the five contiguous GBTDS fields \citep{dustmaps,Zucker2025}. 
Finally, we adopt \(D_{\rm S}=9.0~\mathrm{kpc}\) and \(R_V=2.5\) for the fiducial event, corresponding to the central values of their prior distributions described in Section~\ref{subsec:3_1}, which are approximately summarized by \(D_{\rm S}=9.0\pm1.3~\mathrm{kpc}\) and \(R_V=2.5\pm0.1\). 

Here we adopt the common observable set considered in Section~\ref{subsec:3_1}, in which the three-band lens fluxes and \(\boldsymbol{\mu}_{\rm rel,H}\) are measured from high-resolution imaging, \(t_{\rm E}\) is obtained from the microlensing light curve, and no \(\boldsymbol{\pi}_{\rm E}\) measurement is assumed. 
For the adopted \(M_{\rm L}\), \(D_{\rm L}\), and \(D_{\rm S}\), Equation~\eqref{eq:thetaE_constraint} gives \(\pi_{\rm rel}=0.0222~\mathrm{mas}\) and \(\theta_{\rm E}=0.301~\mathrm{mas}\). 
Using the adopted \(M_{\rm L}\), \([{\rm M/H}]\), \(\tau\), \(D_{\rm L}\), \(A_V\), and \(R_V\), Equations~\eqref{eq:color_F062_F087}--\eqref{eq:mF213_forward} yield \(c_{F062-F087}=3.06~\mathrm{mag}\), \(c_{F087-F213}=3.23~\mathrm{mag}\), and \(m_{F213}=20.61~\mathrm{mag}\). 
We adopt photometric uncertainties of \(\sigma_{F062-F087,\mathrm{obs}}=0.05~\mathrm{mag}\), \(\sigma_{F087-F213,\mathrm{obs}}=0.04~\mathrm{mag}\), and \(\sigma_{m_{F213}}=0.03~\mathrm{mag}\), with the exposure times required to achieve these precisions discussed in Section~\ref{subsec:4_2}. 
To account for intrinsic astrophysical scatter indicated by empirical CMD position--metallicity calibrations, likely at the \(\lesssim0.1~\mathrm{dex}\) level \citep{Duque-Arribas2023}, we adopt \(\sigma_{F062-F087,\mathrm{int}}=0.05~\mathrm{mag}\) and \(\sigma_{F087-F213,\mathrm{int}}=0.03~\mathrm{mag}\), guided by the color shifts associated with metallicity variations in the PARSEC grid. 
These intrinsic-scatter terms are added in quadrature to the photometric uncertainties according to Equations~\eqref{eq:sigma_total_1}--\eqref{eq:sigma_total_2}, yielding total color uncertainties of \(\sigma_{F062-F087,\rm tot}=0.07~\mathrm{mag}\) and \(\sigma_{F087-F213,\rm tot}=0.05~\mathrm{mag}\). 

Because the lens and source velocities enter the likelihood only through \(\boldsymbol{\mu}_{\rm rel,H}\), we directly adopt \(\boldsymbol{\mu}_{\rm rel,H}=(\mu_{{\rm rel,H,E}},\mu_{{\rm rel,H,N}})=(3.0,4.0)~\mathrm{mas\,yr^{-1}}\), where \(\rm E\) and \(\rm N\) denote the east and north components, respectively. 
This corresponds to a magnitude of \(\mu_{\rm rel,H}=5~\mathrm{mas\,yr^{-1}}\), roughly the mean value measured for existing planetary microlensing events \citep{Gould2022_murel}. 
We assign a fractional \(1\sigma\) uncertainty of \(10\%\) to each component and neglect the covariance between the two components. 
Equations~\eqref{eq:murel_Roman} and \eqref{eq:tE_forward} then yield \(t_{\rm E}=22.0~\mathrm{day}\), for which we adopt a \(5\%\) uncertainty. 
These proper-motion and timescale precisions are expected to be achievable for a substantial fraction of Roman events \citep{paper4_Terry2026}. 

The mock measurements for the fiducial event are therefore summarized as
\begin{equation}
\begin{gathered}
\hat{c}_{F062-F087}
=3.06\pm0.07~\mathrm{mag},\\
\hat{c}_{F087-F213}
=3.23\pm0.05~\mathrm{mag},\\
\hat{m}_{F213}
=20.61\pm0.03~\mathrm{mag},\\
\hat{\boldsymbol{\mu}}_{\rm rel,H}
=(3.0\pm0.3,\,4.0\pm0.4)~
\mathrm{mas\,yr^{-1}},\\
\hat{t}_{\rm E}
=22.0\pm1.1~
\mathrm{day}, 
\end{gathered}
\label{eq:fiducial_observables}
\end{equation}
where the quoted color uncertainties include both photometric uncertainties and intrinsic astrophysical scatter around the CMD position--metallicity relation. 
These mock measurements and their uncertainties, together with the adopted prior distributions for \(D_{\rm S}\) and \(R_V\), constitute the inputs to \texttt{MetalLens} for the recovery. 

\begin{figure*}
    \centering
    \includegraphics[width=0.7\textwidth]{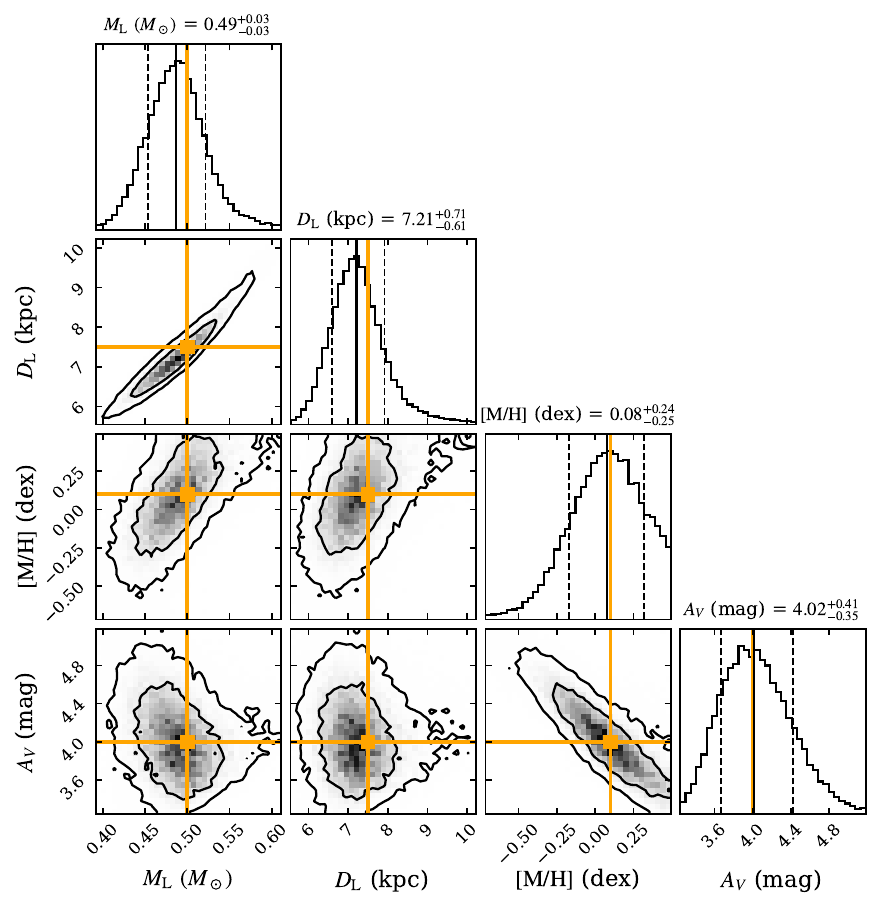}
    \caption{Joint and marginalized posterior distributions of the lens mass \(M_{\rm L}\), lens distance \(D_{\rm L}\), lens metallicity \([{\rm M/H}]\), and extinction to the lens \(A_V\), inferred with \texttt{MetalLens} from the mock observables of the fiducial event. 
    These mock observables consist of the three-band lens photometry, the heliocentric lens--source relative proper motion \(\boldsymbol{\mu}_{\rm rel,H}\), and the event timescale \(t_{\rm E}\), while the inference additionally accounts for source-distance and reddening-vector-direction uncertainties through their prior distributions and for intrinsic scatter around the CMD position--metallicity relation. 
    The orange lines and crosses mark the injected values. 
    The contours in the joint posterior panels enclose \(68\%\) and \(95\%\) of the posterior probability. 
    The solid and dashed black lines in the marginalized posterior panels mark the medians and the 16th and 84th percentiles, respectively, with the corresponding values reported above each panel. 
    Broad uniform priors are assigned to all four parameters, yet the recovery yields well-constrained posterior distributions around the injected values. 
    The correlations among the recovered parameters and their origins are discussed in Section~\ref{subsec:3_3}. 
}
    \label{fig:corner_plot_Mdwarf_lens}
\end{figure*}

\begin{table}
\centering
\caption{Injected and Recovered Parameters for the Fiducial Mock Event}
\label{tab:mock_recovery}
\renewcommand\arraystretch{1.4}
\begin{tabular*}{\columnwidth}{
@{\hspace{\tabcolsep}\extracolsep{\fill}}
lcc
@{\hspace{\tabcolsep}}
}
\hline
Parameter & Injected & Recovered \\
\hline
\(M_{\rm L}\ (M_\odot)\)
& \(0.50\)
& \(0.49^{+0.03}_{-0.03}\) \\

\(D_{\rm L}\ (\mathrm{kpc})\)
& \(7.50\)
& \(7.21^{+0.71}_{-0.61}\) \\

\([{\rm M/H}]\ (\mathrm{dex})\)
& \(+0.10\)
& \(+0.08^{+0.24}_{-0.25}\) \\

\(A_V\ (\mathrm{mag})\)
& \(4.00\)
& \(4.02^{+0.41}_{-0.35}\) \\
\hline
\end{tabular*}
\tablecomments{
The recovered values are posterior medians, with uncertainties given by the 16th and 84th percentiles. 
}
\end{table}

\subsection{Parameter recovery for the fiducial event}
\label{subsec:3_3}

This subsection presents the recovery of the physical parameters of the fiducial mock event.
The target parameters are \(M_{\rm L}\), \(D_{\rm L}\), \([{\rm M/H}]\), and \(A_V\), for which the photometric and microlensing observables provide direct constraints. 
Figure~\ref{fig:corner_plot_Mdwarf_lens} shows their joint and marginalized posterior distributions, while Table~\ref{tab:mock_recovery} compares their injected and recovered values. 
Importantly, all four parameters are assigned broad uniform priors, yet the recovery yields well-constrained posterior distributions around the corresponding injected values. 
The uncertainties in the recovered parameters result from the joint propagation of the photometric uncertainties, the intrinsic astrophysical scatter around the CMD position--metallicity relation, the measurement uncertainties in \(\boldsymbol{\mu}_{\rm rel,H}\) and \(t_{\rm E}\), and the source-distance and reddening-vector-direction uncertainties represented by their prior distributions. 
In particular, the recovered lens metallicity is \([{\rm M/H}]=+0.08^{+0.24}_{-0.25}~\mathrm{dex}\), corresponding to a \(1\sigma\) uncertainty of approximately \(0.25~\mathrm{dex}\). 

Various correlations appear in the joint posterior, including a strong positive \(M_{\rm L}\)--\(D_{\rm L}\) correlation, a strong negative \([{\rm M/H}]\)--\(A_V\) correlation, and weaker positive correlations of \([{\rm M/H}]\) with \(M_{\rm L}\) and \(D_{\rm L}\). 
The origins of these correlations are discussed below. 

The positive \(M_{\rm L}\)--\(D_{\rm L}\) correlation is driven primarily by the lens-flux constraint of Equation~\eqref{eq:lens_flux_constraint}. 
At larger \(D_{\rm L}\), reproducing the same apparent magnitude requires a brighter absolute magnitude and hence a larger \(M_{\rm L}\). 
The \(\theta_{\rm E}\) constraint of Equation~\eqref{eq:thetaE_constraint} provides a second \(M_{\rm L}\)--\(D_{\rm L}\) relation that, together with the lens-flux relation, substantially constrains both parameters. 
However, because it is \(\pi_{\rm rel}=\mathrm{au}(D_{\rm L}^{-1}-D_{\rm S}^{-1})\), rather than \(D_{\rm L}\) alone, that enters the \(\theta_{\rm E}\) constraint, the relatively broad prior on \(D_{\rm S}\) permits a wider range of \(D_{\rm L}\) and hence \(M_{\rm L}\), thereby partially retaining the lens-flux-driven correlation in their joint posterior. 

The negative \([{\rm M/H}]\)--\(A_V\) correlation arises from the color--color dereddening illustrated in Figure~\ref{fig:color_color_plot}. 
For a fixed reddening-vector direction, photometric uncertainties shift the observed color--color position of the reddened lens, causing the corresponding dereddening path to shift parallel to itself and intersect a different part of the intrinsic M-dwarf locus. 
Given the relative orientations of the reddening vector and the locus, an intersection at lower metallicity lies farther from the observed position and therefore corresponds to a larger \(A_V\). 
For a fixed observed color--color position, uncertainty in reddening-vector direction instead rotates the dereddening path, shifting the intersection along the locus and again associating lower \([{\rm M/H}]\) with larger \(A_V\). 

The positive correlations of \([{\rm M/H}]\) with \(M_{\rm L}\) and \(D_{\rm L}\) arise from the dependence of the color--color loci on \(M_{F213}\), as shown in Figure~\ref{fig:color_color_plot}. 
A larger \(M_{\rm L}\) corresponds to a brighter \(M_{F213}\), at which the loci shift toward bluer colors. 
For a fixed dereddened color--color position, this shift leads to a higher inferred metallicity. 
The positive \([{\rm M/H}]\)--\(D_{\rm L}\) correlation follows largely through the positive \(M_{\rm L}\)--\(D_{\rm L}\) correlation. 




\section{Discussion}\label{sec:4_discussion}





In this section, we discuss practical considerations for applying the photometric--microlensing metallicity method, including the chemical interpretation of the inferred metallicity, empirical calibration of the CMD position--metallicity mapping, applicable lenses and observational requirements, parent-sample construction for occurrence--metallicity analysis, effects of unresolved lens companions, and spectroscopic validation for nearby lenses. 
Table~\ref{tab:discussion_roadmap} summarizes their key points, with detailed discussion given in the corresponding sections.

\setlength{\tabcolsep}{4pt}

\newlength{\RoadmapTopicWidth}
\newlength{\RoadmapPointWidth}
\newlength{\RoadmapLocationWidth}

\setlength{\RoadmapTopicWidth}{0.175\textwidth}
\setlength{\RoadmapLocationWidth}{0.065\textwidth}
\setlength{\RoadmapPointWidth}{%
  \dimexpr\textwidth
  -\RoadmapTopicWidth
  -\RoadmapLocationWidth
  -6\tabcolsep-2pt\relax
}

\newcommand{\RoadmapTopic}[1]{%
  \begin{minipage}[t]{\RoadmapTopicWidth}%
    #1%
  \end{minipage}%
}

\newcommand{\RoadmapPoint}[1]{%
  \begin{minipage}[t]{\RoadmapPointWidth}%
    #1%
  \end{minipage}%
}

\newcommand{\RoadmapLocation}[1]{%
  \begin{minipage}[t]{\RoadmapLocationWidth}%
    \centering
    #1%
  \end{minipage}%
}

\newcommand{\RoadmapGap}{%
  \noalign{\vskip 1.30em}%
}

\begin{table*}
\centering
\caption{Practical considerations for applying the photometric--microlensing metallicity method}
\label{tab:discussion_roadmap}

\footnotesize
\setlength{\tabcolsep}{4pt}
\renewcommand{\arraystretch}{1.12}

\begin{tabular}{lll}
\hline
\noalign{\vskip 0.35em}

\RoadmapTopic{\textbf{Topic}} &
\RoadmapPoint{\textbf{Key point}} &
\RoadmapLocation{\textbf{Section}} \\

\noalign{\vskip 0.35em}
\hline
\noalign{\vskip 0.65em}

\RoadmapTopic{Chemical interpretation} &
\RoadmapPoint{%
M-dwarf optical--NIR CMD positions respond to multiple elements, but can still serve as \([{\rm Fe/H}]\) or \([{\rm M/H}]\) indicators because abundance ratios vary in correlated ways within a given Galactic population.
} &
\RoadmapLocation{Sec.~\ref{subsec:4_1}} \\

\RoadmapGap

\RoadmapTopic{Empirical calibration} &
\RoadmapPoint{%
The Roman-filter CMD position--metallicity mapping should be empirically calibrated using FGK+M binaries or clusters.
Calibrators should follow the abundance trends of the target lens population, e.g., local thick-disk stars for metal-poor bulge lenses due to their similar \(\alpha\)-enhancement trends.
} &
\RoadmapLocation{Sec.~\ref{subsec:4_1}} \\

\RoadmapGap

\RoadmapTopic{Reddening-vector direction} &
\RoadmapPoint{%
The reddening-vector direction should be empirically calibrated, using bulge red-clump photometry for sightline-specific calibration or grism spectroscopy of field stars for distance-resolved calibration.
} &
\RoadmapLocation{Sec.~\ref{subsec:2_4}} \\

\RoadmapGap

\RoadmapTopic{Applicable sample and\\ observation} &
\RoadmapPoint{%
\textbf{Sample:} $\sim0.4$--$0.6\,M_\odot$ lenses out to bulge distances with sufficient late-time lens--source separation, corresponding to $\sim150$ planetary systems under the adopted Roman yield forecast.
\textbf{Observation:} $\sim8~\mathrm{hr}$ of $F062$ imaging per field roughly a decade after Roman launch ($\sim40~\mathrm{hr}$ in total across the five contiguous GBTDS fields).
} &
\RoadmapLocation{Sec.~\ref{subsec:4_2}} \\

\RoadmapGap

\RoadmapTopic{Parent lens sample} &
\RoadmapPoint{%
Apply the same metallicity inference homogeneously to events with and without detected planets to construct the parent sample required for occurrence--metallicity analysis.
} &
\RoadmapLocation{Sec.~\ref{subsec:4_3}} \\

\RoadmapGap

\RoadmapTopic{Unresolved companions} &
\RoadmapPoint{%
\textbf{Individual lens:} unresolved companions bias inferred metallicity upward.
\textbf{Population statistics:} the small and similar hidden-companion fractions in both planet hosts and the parent sample limit the impact on occurrence analysis, while the one-sided bias cannot make planet hosts appear artificially metal-poor.
} &
\RoadmapLocation{Sec.~\ref{subsec:4_4}} \\

\RoadmapGap

\RoadmapTopic{Spectroscopic validation} &
\RoadmapPoint{%
Nearby-lens metallicities can be cross-validated spectroscopically.
A \(\sim3~\mathrm{hr}\) Keck/OSIRIS IFS exposure enables validation for a \(0.6\,M_\odot\) lens out to \(\sim3\) kpc or a \(0.4\,M_\odot\) lens out to \(\sim1.7\) kpc.
} &
\RoadmapLocation{Sec.~\ref{subsec:4_5}} \\

\noalign{\vskip 0.65em}
\hline
\end{tabular}
\end{table*}

\subsection{Chemical interpretation and empirical calibration} 
\label{subsec:4_1}

The metallicity label \([{\rm M/H}]\) used in this work should be interpreted as the scaled-solar \([{\rm M/H}]\) required to reproduce the lens CMD position. 
The reason is that we adopt PARSEC isochrones, which assume a scaled-solar abundance mixture, to provide the CMD position--metallicity mapping. 
This is adequate for the present feasibility demonstration, but real applications will require an empirical calibration of the Roman-filter CMD position--metallicity mapping to account for theoretical-isochrone systematics and non-scaled-solar abundance patterns. 

A useful empirical calibration must tie the observable to a chemical-abundance label that the observable can meaningfully constrain. 
For M dwarfs, the choice of this label is subtle because the relevant observables are shaped by molecular opacity and pseudo-continuum placement rather than by isolated atomic lines of individual elements, as in classical FGK-star spectroscopy. 
In M-dwarf photometric metallicity methods that this work extends, the optical--NIR color displacement at fixed near-infrared absolute magnitude is driven mainly by Ti- and O-bearing molecular opacity, especially TiO, but it is not merely a measurement of Ti or O abundances. 
The carbon abundance also matters because CO is energetically favored and can lock up a substantial fraction of the available oxygen, thereby reducing the oxygen left to form TiO. 
Separately varying the Ti, O, or C abundance can produce comparable changes in the optical pseudo-continuum level, with a 0.2 dex abundance change shifting this level by up to \(\sim 40\%\) for Ti and O and \(\sim 20\%\) for C at \(T_{\rm eff}=3000\,{\rm K}\) \citep{Rains2024}. 
As a result, individual elemental abundances can substantially reshape the pseudo-continuum through their effects on dominant molecular absorbers, making the resulting color displacement degenerate with respect to different abundance combinations. 
This issue is not unique to photometric metallicity methods. 
Although direct FGK-style abundance analysis remains difficult for M dwarfs, empirical spectroscopic metallicity methods have been developed using optical or near-infrared spectra at moderate to high resolution. 
These methods measure equivalent widths of metal-sensitive lines or blended features relative to the pseudo-continuum, usually together with temperature-sensitive molecular indices, and calibrate these observables with FGK+M binaries \citep[e.g.,][]{Rojas-Ayala2010,Rojas-Ayala2012,Terrien2012,Mann2013,Newton2014,Neves2014,Veyette2017,Rains2024}. 
These methods can reach scatters \(\sim 0.1\) dex, but their observables still reflect degenerate abundance combinations rather than individual elemental abundances. 
In near-infrared spectroscopic methods, for example, Na and Ca features are measured relative to a pseudo-continuum set largely by H\(_2\)O absorption, whose strength depends on how much oxygen remains after CO formation. 
Consequently, independently varying C and O abundances over the \(\pm 0.2\) dex range can make the inferred metallicity span more than 1 dex \citep{Veyette2016}. 
Moreover, beyond the fact that both photometric and spectroscopic methods do not directly trace individual elemental abundances, most of these empirical relations are calibrated to \([{\rm Fe/H}]\), even though their observables are not based on \text{Fe\,\textsc{i}} absorption lines and therefore do not directly probe the Fe abundance \citep{Veyette2016,Veyette2017,Rains2024}. 

Crucially, these chemical degeneracies and the indirect connection to Fe do not invalidate empirical \([{\rm Fe/H}]\) calibrations. 
They work because abundance ratios vary in correlated ways rather than independently. 
In particular, ratios that are relevant to M-dwarf molecular opacity, such as \([{\rm Ti/Fe}]\) and the C/O balance encoded by \([{\rm C/Fe}]\) and \([{\rm O/Fe}]\), follow relatively tight trends with \([{\rm Fe/H}]\) within a given Galactic population, such as the thin disk or the thick disk, as a result of Galactic chemical evolution \citep{Bensby2014_Disk,Nissen2014_CFe_OFe_CO,Veyette2016,Bensby2017,Veyette2017,Rains2021}. 
%
Thus, an observable that does not directly probe Fe can still serve as a statistical \([{\rm Fe/H}]\) indicator when calibrated against benchmarks with known \([{\rm Fe/H}]\). 
More generally, the metallicity label inferred by an M-dwarf metallicity method is set by the abundance label used for its calibration, most commonly \([{\rm Fe/H}]\) or \([{\rm M/H}]\). 
Population-dependent differences in these abundance trends are discussed below. 

A practical Roman application will require an empirical calibration of the Roman-filter CMD position--metallicity relation, for which three routes are available: (i) synthetic Roman photometry of existing nearby FGK+M binary calibrators, (ii) direct Roman observations of more distant FGK+M binaries, and (iii) cluster-based calibration. 
For route (i), Roman magnitudes can be synthesized from flux-calibrated spectra for existing FGK+M binary calibrators. 
The reason is that these widely used calibrators are overwhelmingly local, with nearly all systems within \(100~\mathrm{pc}\) \citep{Mann2013,Montes2018}. 
At these distances, the M-dwarf photometry needed for the calibration would generally saturate even in the shortest Roman exposures. 
Instead, the M dwarfs in these nearby binary calibrators can be placed on the Roman-filter CMD through synthetic Roman photometry, using Gaia XP spectra for \(F062\) and \(F087\) and SPHEREx spectra for \(F213\) \citep{galaxy_in_your_preferred_colours, SPHEREx}, while retaining the metallicity labels inherited from their FGK primaries. 
For route (ii), Roman can directly observe more distant, and therefore fainter, M dwarfs in FGK+M binaries at distances of several hundred parsecs or more, selected from Gaia spatially resolved binary catalogs \citep[e.g.,][]{El-Badry2021_Gaia_binary_catalog}. 
This would extend the binary calibration into the unsaturated Roman-photometry regime, while the metallicities of the FGK primaries can be supplied by spectroscopic surveys \citep{Qiu2024}. 
In practice, these calibrators should be selected from low-reddening sight lines, preferably at high Galactic latitude. 
For route (iii), open and globular clusters spanning a broad metallicity range can serve as calibrators \citep{Rains2024}. 
Because cluster members are chemically homogeneous and their metallicities can be measured from brighter members, Roman photometry of the cluster M dwarfs would trace a lower-main-sequence locus in the Roman-filter CMD at a known metallicity. 
A set of clusters with different metallicities would therefore provide empirical CMD loci across metallicity, directly calibrating the Roman-filter CMD position--metallicity mapping. 
In practice, cluster calibrators should be restricted to systems with low or well-characterized mean and differential reddening. 


The calibrators and Roman microlensing lenses will not necessarily sample identical Galactic environments, so population-dependent abundance trends should be accounted for in the calibration. 
This is particularly important at subsolar metallicity, where the local thin- and thick-disk sequences differ in \(\alpha\)-enhancement, while microlensed bulge dwarfs largely follow the \(\alpha\)-enhanced thick-disk sequence \citep[see Fig.~20 of][]{Bensby2017}. 
For bulge lenses, the metal-poor end of the calibration should therefore preferentially be constructed from \(\alpha\)-enhanced, thick-disk-like calibrators. 
Residual population-dependent differences can be treated as a calibration systematic.




\subsection{Applicable lenses and observational requirements}
\label{subsec:4_2}

The applicable lens-mass range of the photometric--microlensing metallicity method is bounded at the high-mass end by the transition out of the M-dwarf regime and at the low-mass end by lens-flux detectability. 
The high-mass boundary arises because the cool atmospheres of M dwarfs allow abundant molecules to form, and the resulting broad absorption bands imprint a strong metallicity dependence on optical broad-band fluxes and hence on the positions of M dwarfs in the optical--NIR CMD. 
For warmer FGK stars, the same optical--NIR color provides substantially weaker metallicity leverage, as illustrated in Figure~\ref{fig:cmd}.\footnote{Classical broad-band photometric metallicities for FGK stars instead exploit line blanketing at substantially bluer wavelengths than those used here, typically requiring photometry that includes a band such as SDSS \(u\); for example, \(g-r\) primarily traces effective temperature, while \(u-g\) provides metallicity sensitivity at fixed \(g-r\) \citep{Ivezic2008}. 
However, extending this low-reddening approach to highly reddened microlensing lenses with an analogous color--color dereddening would be less well conditioned than in the M-dwarf case. 
This is because, at the expected level of uncertainty in the lens absolute magnitude constrained by microlensing observables, the different-metallicity FGK loci would be less cleanly separated in the \(ugr\) color--color plane than in the M-dwarf optical--NIR case shown in Figure~\ref{fig:color_color_plot}, so the intersection of the reddening vector with stellar loci would remain consistent with a broader range of metallicities. 
}
The method is therefore limited to M dwarfs, corresponding to an approximate upper mass limit of \(0.6\,M_\odot\). 

At the low-mass end, because the method is designed to be applicable to highly reddened lenses out to bulge distances, the practical boundary is set mainly by the lens-photometry depth achievable at the precision needed for useful metallicity inference. 
The Roman implementation considered here uses three-band lens photometry in \(F062\), \(F087\), and \(F213\). 
A band as blue as \(F062\) is required because it samples the optical wavelength range in which M-dwarf molecular opacity imprints the strongest metallicity dependence on broad-band fluxes.\footnote{Roman \(F062\) imaging could be replaced by observations from another facility with similar wavelength coverage, sufficient photometric depth, angular resolution comparable to that of Roman \(F062\), and a wide field of view; CSST \(r\)-band imaging is a promising alternative.} 
Implementing the method using only \(F087\) and redder filters yields a stellar locus that spans substantially smaller ranges in color--color space over a given metallicity interval, and the metal-poor part of the resulting locus is nearly parallel to the reddening vector; together, these effects make the inference much more vulnerable to photometric errors and uncertainty in the reddening-vector direction. 
Under the current GBTDS design, observations in \(F087\) and \(F213\) are planned at six-hour cadence during the high-cadence seasons, yielding many epochs that can be stacked to substantial depth, whereas \(F062\) imaging is limited to short photometric snapshots \citep{ROTAC_final_report}. 
These \(F062\) snapshots are too shallow to reach the \(F062-F087\) color precision adopted for the fiducial bulge M-dwarf lens in Section~\ref{subsec:3_2}, so additional deep \(F062\) imaging is required. 

The depth of this additional \(F062\) imaging therefore largely sets the practical lower-mass reach of the method. 
For the fiducial \(0.5\,M_\odot\) lens at \(D_{\rm L}=7.5~\mathrm{kpc}\) with \(A_V=4~\mathrm{mag}\) adopted in Section~\ref{subsec:3_2}, which has \(m_{F062}\simeq27~\mathrm{mag}\), calculations with the \href{https://roman.etc.stsci.edu/}{Roman WFI Exposure Time Calculator} indicate that reaching \(S/N\simeq25\) in \(F062\), a target chosen to support the \(0.05~\mathrm{mag}\) uncertainty adopted for \(F062-F087\) in the mock recovery, requires \(\sim8~\mathrm{hr}\) of exposure with a near-optimal aperture-extraction strategy under the background conditions for the GBTDS field coordinates and representative bulge-season dates. 
In practice, the individual lens and source fluxes will be measured through two-star point-spread-function (PSF) fitting of the dithered images, with the additional uncertainty in separating the lens and source fluxes discussed below. 
With distance and extinction held fixed, the same \(\sim8~\mathrm{hr}\) exposure would yield \(S/N\sim11\) for a \(0.4\,M_\odot\) lens, still permitting a useful metallicity constraint, though at somewhat reduced precision. 
The applicable mass range for bulge-distance lenses at this photometric depth is therefore approximately \(0.4\)--\(0.6\,M_\odot\), while lower-mass lenses remain accessible when they are closer or less extinguished. 
Extending this range below \(0.4\,M_\odot\) at bulge distances would require rapidly increasing \(F062\) exposure times while yielding only a modest expected increase in the number of lenses with useful metallicity constraints, making the adopted depth a practical compromise between exposure time and sample size.

In addition to the photometric depth discussed above, the precision attainable for the lens flux also depends on the lens--source separation at the follow-up epoch. 
When the lens and source remain only partially resolved in high-resolution follow-up images, which is the case for many existing microlensing lens-flux measurements, their individual fluxes must be inferred jointly through two-star PSF fitting. 
Existing high-resolution lens-flux measurements obtained with the Hubble Space Telescope (HST) at lens--source separations below one PSF full width at half maximum (FWHM) commonly yield enlarged uncertainties in the individual lens and source fluxes compared with the uncertainty in their combined flux, because the lens and source can trade flux in the two-star fit; this additional uncertainty generally decreases as the lens--source separation increases relative to the PSF FWHM \citep{HST_Bennett2015_OGLE-2005-BLG-169, HST_Bhattacharya_OGLE-2012-BLG-0950, HST_Bennett2024_MOA-2008-BLG-379, HST_Terry2024_MOA-2007-BLG-192, HST_Bennett2024_OGLE-2012-BLG-0563, HST_DexBhadra2026_KMT-2019-BLG-0253}. 
Roman provides an advantage in \(F087\) and \(F213\), because the six-hour-cadence microlensing light curves independently constrain the source fluxes in these bands, thereby restricting flux trading between the lens and source in the corresponding two-star fits. 
Although \(F062\) lacks such a light curve and therefore an equivalent source-flux constraint, it has Roman's sharpest PSF, with an FWHM of approximately \(58~\mathrm{mas}\), as tabulated in the \href{https://github.com/RomanSpaceTelescope/roman-technical-information/blob/v1.3/data/WideFieldInstrument/Imaging/FiltersSummary/filter_parameters.ecsv}{Roman WFI Filter Parameters Table}. 
For events with \(\mu_{\rm rel}\gtrsim5~\mathrm{mas\,yr^{-1}}\), approximately the median value for existing planetary microlensing events \citep{Gould2022_murel}, baselines of roughly a decade yield lens--source separations of \(\gtrsim50~\mathrm{mas}\), comparable to or exceeding the \(F062\) PSF FWHM. 
At such separations, the additional uncertainty arising from lens--source flux trading should be reduced, though not necessarily eliminated, provided that sufficient subpixel dithering is used to mitigate undersampling of the \(F062\) PSF. 
These considerations favor obtaining the additional \(\sim8~\mathrm{hr}\) of sufficiently dithered \(F062\) imaging per field in a late-time campaign roughly a decade after Roman launch, corresponding to a total of \(\sim40~\mathrm{hr}\) to cover all five contiguous GBTDS fields, with a modestly higher \(\mu_{\rm rel}\) threshold applied to events from the later GBTDS seasons to compensate for their shorter baselines. 
Alternatively, distributing a comparable total \(F062\) integration across the GBTDS seasons to obtain low-cadence \(F062\) microlensing light curves could similarly constrain the \(F062\) source flux, potentially enabling earlier application of the method. 
The exposure-time requirement and timing of the late-time campaign should ultimately be refined through image-level simulations. 


Once these observational requirements are met, the method should be applicable to a non-negligible fraction of Roman microlensing lenses. 
Because the adopted photometric depth is designed to reach lenses out to bulge distances, the principal selection criteria are whether the lens lies within the applicable mass range and whether the lens and source are sufficiently separated at the late-time \(F062\) imaging epoch. 
\citet{Penny2019} predict that Roman will detect approximately \(1400\) bound planets through microlensing, of which roughly one quarter orbit hosts with masses within the \(0.4\)--\(0.6\,M_\odot\) range over which the method is applicable, as indicated by the \href{https://github.com/mtpenny/wfirst-ml-figures/blob/master/sensitivity/c62cassan.160.47s.layout_7f_3_covfac_2.81_432.0_1.0.sample0} {publicly available simulation output} accompanying that study. 
To estimate the fraction satisfying the lens--source separation condition, we use the empirical planetary-event \(\mu_{\rm rel}\) distribution of \citet{Gould2022_murel}, rather than the simulated \(\mu_{\rm rel}\) values of \citet{Penny2019}, which the authors note are systematically overestimated. 
At a representative late-time \(F062\) imaging epoch around a decade after launch, and accounting for the season-dependent baselines, this distribution suggests that somewhat less than half of the mass-selected systems could reach lens--source separations comparable to those discussed above, with the precise fraction depending sensitively on the campaign timing. 
Combining the mass and separation selections suggests that around \(150\) planetary systems could be accessible to the method under the \citet{Penny2019} yield forecast, with the uncertainty in this estimate depending on the lens--source deblending performance ultimately achieved.

The method could also be applied to microlensing events previously discovered by ground-based surveys, many of which should already have sufficiently large lens--source separations for high-resolution multiband follow-up and could therefore provide an earlier opportunity to apply the method. Although these historical events are distributed over a much wider area of the sky than the compact GBTDS fields, this loss of survey efficiency is partially offset by the lower extinction toward many ground-based microlensing fields, which can substantially reduce the optical exposure time required for a given photometric precision.

\subsection{A homogeneously constructed parent lens sample for occurrence--metallicity measurements}
\label{subsec:4_3}

Measuring planet occurrence as a function of host metallicity requires not only metallicities for the detected planet hosts, but also metallicities for a parent lens sample that defines the denominator of the occurrence calculation. 
Here the parent lens sample refers to microlensing lenses that satisfy the same M-dwarf lens selection and metallicity-inference requirements, regardless of whether a planet is detected. 
This is particularly important for microlensing, because the lens population spans a wide range of Galactic environments, from the nearby disk to the inner disk and bulge, where the underlying metallicity distribution varies substantially \citep{Hayden2015, Zoccali2017_bulge_metallicity}. 
External metallicity maps do not by themselves provide the occurrence denominator, which must be tied to the event-level planet-detection efficiencies of the microlensing sample. 
A homogeneous metallicity inference across the full parent lens sample is therefore essential.

The same Roman GBTDS data provide a natural way to construct this parent lens sample, by applying our photometric--microlensing metallicity inference homogeneously to microlensing events with and without planet detections. 
The method described in this work does not rely on the presence of a planet, but only lens fluxes in the relevant bands and microlensing constraints on the lens mass--distance relation. 
It can therefore be applied not only to planet-hosting events, but also to the much larger sample of single-lens events without detected planets, which are expected to outnumber planetary events by more than an order of magnitude \citep{Penny2019}. 
For M-dwarf lenses in single-lens events, combining their multi-band lens fluxes with microlensing observables, especially \(\theta_{\rm E}\) inferred from \(\theta_{\rm E}=t_{\rm E}\mu_{\rm rel}\), where \(t_{\rm E}\) is measured from the microlensing light curve and \(\mu_{\rm rel}\) is obtained from the Roman multi-epoch high-resolution imaging, allows the same metallicity inference as for planet hosts. 
The union of these single-lens events and the planet-hosting events therefore forms the homogeneously analyzed parent lens sample for the occurrence calculation. 
As a by-product, this parent sample will also provide a large set of M-dwarf metallicity measurements across diverse Galactic environments. 
%
%

Importantly, an event without a detected planet does not imply a planet-free lens, because a planet of the relevant class may simply have escaped detection. 
Instead, each event in the parent lens sample, whether or not a planet is detected, contributes to the occurrence denominator according to the probability that a planet of that class would have been detected in its light curve. 
For a planet class of interest and a metallicity bin \(Z_k\), the occurrence rate can be written schematically as
\begin{equation}
f_{\rm p}(Z_k) \simeq
\frac{N_{\rm detected}(Z_k)}
{\sum_{i\in Z_k}\epsilon_i},
\end{equation}
where \(N_{\rm detected}(Z_k)\) is the number of detected planets of that class in the metallicity bin \(Z_k\), \(\epsilon_i\) is the detection efficiency of event \(i\) to the relevant planet class, and the sum runs over all microlensing events in the parent lens sample whose lens metallicities lie in \(Z_k\), including both events with and without detected planets. 
In practice, the hard bin assignment can be replaced by metallicity-posterior weights in both the numerator and denominator. 
In this way, the homogeneously inferred metallicities of the parent lens sample can be combined with standard microlensing sensitivity calculations to measure the occurrence--metallicity relation. 

\subsection{Unresolved lens companions}
\label{subsec:4_4}








An important caveat of the method is the possible presence of a luminous companion to the lens that is unresolved in high-resolution imaging and not identified from the microlensing light curve, because interpreting such a system as a single star would bias the inferred lens metallicity upward. 
Consider an equal-mass companion to an M-dwarf lens. 
The flux in each band is then doubled, while the broad-band colors and the microlensing constraints are unchanged. 
As a result, the observed color-color position is unchanged, so the color-color dereddening procedure shown in Figure~\ref{fig:color_color_plot} returns nearly the same intrinsic color and extinction as it would for an isolated M-dwarf lens.
However, a single-star interpretation explains the excess flux as a brighter absolute magnitude, by about \(0.75\) mag in each band. 
Thus, in the intrinsic CMD of Figure~\ref{fig:cmd}, the inferred lens has nearly the same intrinsic color as the isolated M dwarf, but has \(M_{F213}\) brighter by about \(0.75\) mag. 
This moves the inferred CMD position toward more metal-rich isochrones, causing the metallicity to be overestimated. 
Applying the mock-recovery pipeline to this equal-mass binary case gives a metallicity bias of order \(\sim0.4\) dex. 

Near-equal-mass M-dwarf companions are the most problematic case for metallicity. 
A much fainter companion contributes little light and therefore produces only a small perturbation. 
If the companion is more massive but still an M dwarf, the blended flux may instead be dominated by the companion. 
In that case, the single-star solution may describe the brighter companion rather than the lens. 
Although the inferred mass would then be inappropriate for the actual lens, the metallicity bias should be much smaller than in the equal-mass case, because the recovered CMD position is mainly set by the brighter companion and the lens should share the same composition as its companion. 
A substantially hotter companion, such as an FGK dwarf, would usually dominate the optical--NIR spectral energy distribution (SED) and move the blended solution outside the M-dwarf-lens sample considered here. 

Unresolved multiplicity is a common limitation of photo-astrometric stellar-parameter inference \citep{Bailer-Jones2011, Andrae2018_GaiaDR2, Anders2022}, whereas stellar surveys with usable spectra can sometimes identify such systems through inconsistencies between spectral shape and absolute luminosity \citep{LiJiadong2025_Gaia_distinguish_binary_from_spectra}. 
For a near-equal-mass binary, the normalized spectrum is similar to that of either component, but the system is overluminous once the distance is fixed by parallax. 
Fitting such a system as a single star therefore produces a mismatch between the normalized spectral shape and the absolute spectral flux scale. 
For microlensing lenses, the corresponding test would be to compare the effective temperature inferred from the lens spectrum with that inferred from the single-star photometric--microlensing solution. 
Roman will obtain grism spectroscopic snapshots of the GBTDS fields, with wavelength coverage \(1.00{-}1.93~\mu{\rm m}\) and resolving power \(R\simeq460\) \citep{ROTAC_final_report}. 
In principle, such spectra could provide an effective-temperature diagnostic for M dwarfs through the pseudo-continuum curvature imposed by H$_2$O absorption bands in the near-infrared \citep[e.g.,][]{Rojas-Ayala2012, Terrien2012}. 
In practice, however, this will not be a generally usable solution for the bulge M-dwarf lenses targeted here. 
Although the coadded grism exposure may be long enough to reach modest nominal \(S/N\) for bulge M dwarfs, the limiting issue is spectral confusion in the dense bulge fields. 
These lenses are much fainter than the practical confusion-limited sensitivity for Roman slitless spectra, so their dispersed traces will often be strongly contaminated by neighboring stars. 
Moreover, the lens and source spectra will be blended in the same slitless spectral trace. 
Therefore, Roman grism spectra cannot provide a practical event-by-event diagnostic of unresolved companions for the M-dwarf lens sample. 
A small subset of events with both precise \(\theta_{\rm E}\) and precise \(\pi_{\rm E}\) could provide an independent lens-mass estimate and thereby identify an overluminous lens, but this will not be available for most bulge lenses. 

Although unresolved companions are an important caveat for individual lens metallicities, they are unlikely to dominate the population-level statistics targeted in this work: the occurrence--metallicity relation for cold low-mass planets, and the possible low-metallicity cutoff for cold low-mass planet formation. 
The most problematic companions are those at intermediate projected separations, roughly \(10{-}100\,{\rm au}\), which are generally too wide to often produce a detectable binary-lens anomaly in microlensing light curve, but too close to be resolved in Roman imaging for bulge lenses. 
Empirical M-dwarf multiplicity statistics imply that only \(\sim 10\%\) M dwarfs have companions at separations of \(10{-}100\,{\rm au}\) \citep[see Figure~2 of][]{Offner2023_binary_review}. 
Because this hidden population is small, its metallicity bias is one-sided, and the occurrence analysis uses a homogeneously analyzed parent lens sample, its impact on the two main science goals is limited in different ways: 

\emph{(i) Occurrence--metallicity relation.}
As discussed in Section~\ref{subsec:4_3}, the occurrence calculation uses a sensitivity-weighted parent lens sample whose metallicities are inferred with the same photometric--microlensing method as for the detected planet hosts. 
This homogeneous treatment is the main reason that unresolved companions are less problematic for the occurrence--metallicity relation than for individual lens metallicities. 
If the unresolved-companion fraction is independent of whether a planet is detected, then in any true-metallicity bin the same fraction of detected planet hosts and parent-sample events is scattered to higher measured metallicity, so the numerator and denominator of the occurrence estimate are distorted in the same way and the effect largely cancels in the occurrence ratio. 
The residual caveat is that stellar companions at tens of au may suppress the occurrence of planets orbiting one component of the binary, so detected planet hosts could have a lower unresolved-companion fraction than the parent sample \citep{Kraus2016_planet_surrounding_one_companion_of_binary_suppress, Moe2021_planet_surrounding_one_companion_of_binary_suppress}. 
Even in this case, any differential effect is limited by the fact that the hidden-companion population is expected to comprise only \(\sim10\%\) of the parent sample, and planet suppression can only reduce this already small fraction among detected planet hosts. 
It does not affect the remaining majority of the sample. 
The associated uncertainty can therefore be quantified by repeating the occurrence analysis over plausible unresolved-companion fractions and planet-occurrence suppression factors in binaries. 
\emph{(ii) Low-metallicity cutoff.} 
The one-sided nature of the bias is especially important for constraining the low-metallicity cutoff for cold low-mass planet formation. 
Unresolved companions can only bias inferred metallicities upward. 
As a result, they cannot make an intrinsically metal-rich planet host appear to have a very low metallicity. 
Thus, if cold low-mass planet hosts are found at very low inferred metallicity, their low-metallicity interpretation would be robust, and any cutoff would be pushed to lower metallicity. 
At most, unresolved companions could make the inferred cutoff slightly more metal-rich by moving a small fraction of intrinsically low-metallicity planet hosts to higher inferred metallicity. 
Given the expected only \(\sim10\%\) contamination fraction, this effect is unlikely to shift the inferred cutoff significantly unless the planet-host sample near the cutoff is very small. 

Taken together, the arguments above indicate that the hidden companions should have a limited impact on the population-level statistics. 
Moreover, this effect can be further corrected statistically in the final analysis. 
A practical implementation is to randomly assign unresolved companions to a plausible fraction of the lens sample, shift the metallicities of the affected systems downward according to the corresponding bias distribution, and repeat the occurrence and cutoff analyses over many Monte-Carlo realizations. 

Companions that remain unresolved in imaging but are detected directly in the microlensing light curve are not part of this hidden-contaminant class. 
In such cases, the light curve constrains the binary mass ratio, so the unresolved lens flux can be modeled as the sum of two stellar components rather than being interpreted as a single star. 
These systems could therefore provide a useful by-product: a sample of binary M dwarfs in the bulge and inner disk with mass and metallicity measurements. 
Such a sample would probe low-mass stellar multiplicity as a function of metallicity in a Galactic regime that is difficult to access with existing surveys. 

\subsection{Spectroscopic validation for nearby disk lenses}
\label{subsec:4_5}

Although our method is logically self-contained, it would be useful to cross-validate it with an independent method applied to the same lenses.
The empirical M-dwarf spectroscopic metallicity methods introduced in Section~\ref{subsec:4_1} provide such an alternative. 
As summarized there, these methods use moderate- to high-resolution optical or near-infrared spectra to measure metal-sensitive lines or blended features relative to the pseudo-continuum, usually together with temperature-sensitive molecular indices, and calibrate these observables with FGK+M binaries, reaching scatters \(\sim 0.1\) dex. 

For microlensing lenses, this validation generally requires diffraction-limited rather than seeing-limited spectroscopy. 
In seeing-limited spectroscopy, the lens spectrum is often blended with the source and unrelated field stars in crowded bulge fields, and the broad PSF forces the use of a large extraction aperture that includes substantial sky background. 
Because microlensing lenses are generally faint, this additional background can significantly degrade the lens-spectrum \(S/N\). 
A practical facility for such validation is the Keck/OSIRIS integral-field spectrograph (IFS), which operates with adaptive optics (AO) and can deliver diffraction-limited near-infrared integral-field spectra at a resolution of \(R\sim3800\), with spaxel scales as fine as \(20\) mas \citep{OSIRIS_paper}. 
By contrast, JWST/NIRSpec has a spectroscopic spatial scale of \(\sim0.1''{-}0.2''\), which is generally insufficient to isolate the lens spectrum from that of the source. 
The AO correction concentrates the lens flux into a compact PSF, reducing the effective background in the extracted spectrum. 
At the same time, the IFS data cube allows the lens spectrum to be extracted by combining an adaptive set of spaxels chosen to maximize \(S/N\) while limiting source and field-star contamination. 
This extraction is most effective once the lens and source are spatially resolved, or in cases where the lens is much brighter than the source. 
The resulting \(R\sim3800\) spectrum can then be binned to \(R\sim2000\), comparable to the resolution used in previous moderate-resolution near-infrared M-dwarf spectroscopic-metallicity calibrations \citep{Rojas-Ayala2010,Rojas-Ayala2012,Terrien2012,Mann2013,Newton2014}, to improve the \(S/N\). 

With this observational setup, the empirical spectroscopic method can be applied to the nearby subset of the Roman M-dwarf lenses that is bright enough to yield spectra with sufficient \(S/N\) for metallicity measurements. 
For \(K_s\lesssim17.5\)~mag lenses, a \(\sim3\) hr OSIRIS exposure is expected to reach \(S/N\gtrsim50\) after binning to \(R\sim2000\), based on the \href{https://oirlab.ucsd.edu/osiris/etc/}{OSIRIS Exposure Time Calculator}, which should make a metallicity measurement with a precision of order \(0.1\)~dex feasible. 
This magnitude limit corresponds roughly to a \(0.6\,M_\odot\) lens out to \(\sim3~\mathrm{kpc}\), or a \(0.4\,M_\odot\) lens out to \(\sim1.7~\mathrm{kpc}\). 
For this subset, comparing the spectroscopic metallicities with those inferred from our method would provide an independent check on our metallicity measurements and reveal any systematic bias in the photometric calibration. 
For more distant M-dwarf lenses, however, spectroscopic follow-up will remain difficult with current 10 m-class facilities. 
Looking ahead, future 30 m-class AO-fed spectrographs should extend such validation to bulge M-dwarf lenses. 
Nevertheless, such spectroscopy would remain expensive targeted follow-up rather than a homogeneous route for the full Roman sample. 
Our photometric--microlensing method therefore remains the practical route to host-metallicity measurements for the much larger bulge and inner-disk Roman M-dwarf lens population. 

\acknowledgments
We thank Subo Dong, Tianjun Gan, Scott Gaudi, Jiadong Li, and Raphael Oliveira for helpful discussions and suggestions. We thank the anonymous referee for helpful comments that improved the manuscript. J.Z., W.Z., H.Y., and S.M. acknowledge support by the National Natural Science Foundation of China (Grant No. 12133005, PI: S.M.).








\bibliography{MetaLens.bib}

\end{CJK*}
\end{document}